# AAS Climate Change Task Force Report

Final Version: 21 May 2024

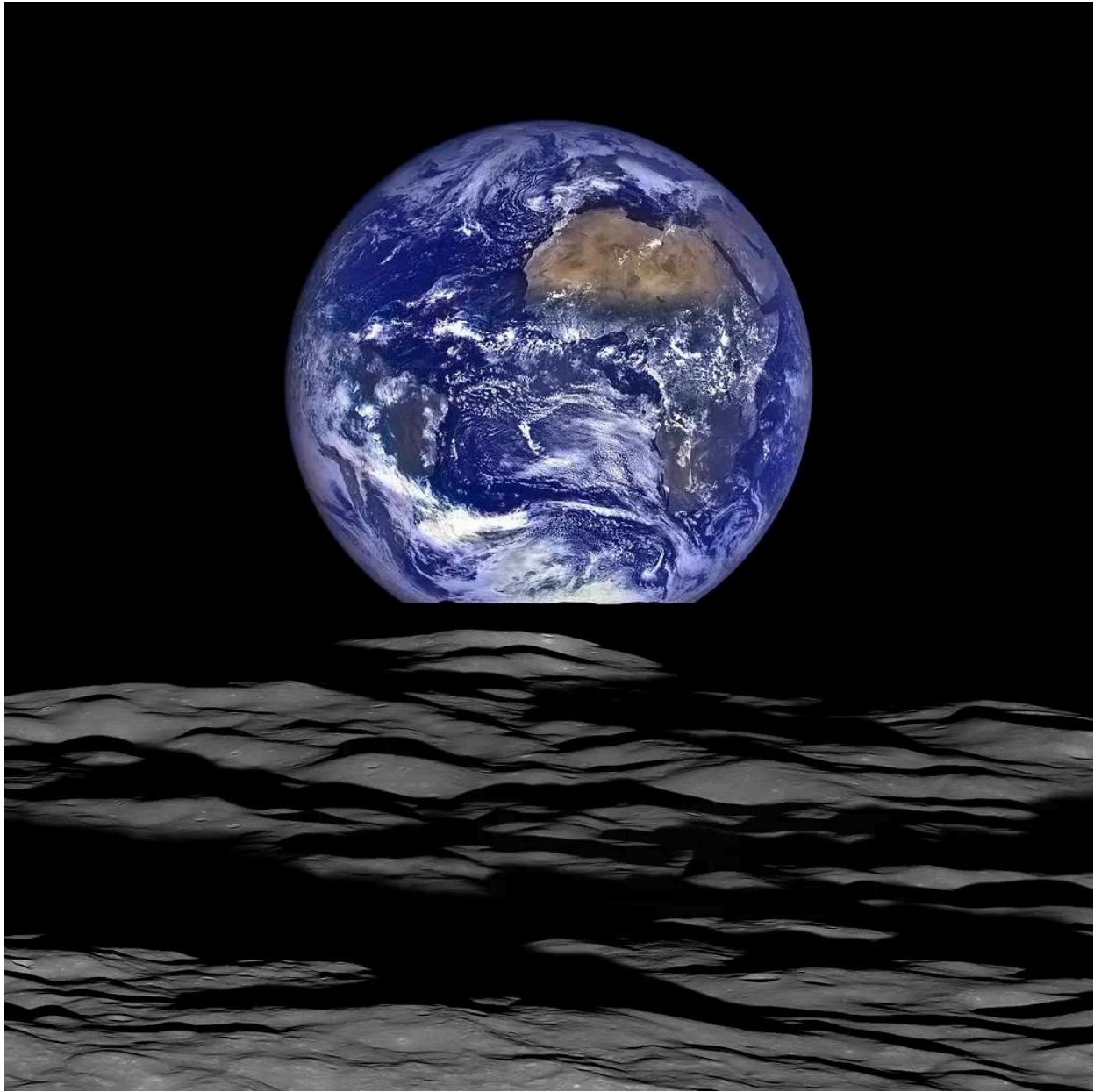

Caption: Earthrise from the moon, captured by the Lunar Reconnaissance Orbiter (Credit: NASA).

"When someone finds out that I am an astronomer, they often want to know if I think there is life beyond Earth. The answer is simple: I don't know.  But I do know we are destroying the only life in the Universe we are certain of." – Anonymous AAS survey comment

## Executive Summary

The AAS Strategic Plan for 2021-26 called for the creation of this task force to identify how the AAS can meet the goals of the Paris Agreement, which are to cut global carbon emissions in half by 2030 and achieve carbon neutrality around 2050.  Our task force consists of fourteen volunteers who were selected to ensure that our recommendations also align with other goals of the AAS as identified in the AAS Mission Statement and the Strategic Plan.

The severity of the problem, and the urgency to act, cannot be overstated.  Simply put, climate change is a crisis.  The bad news is that it is already causing severe harm.  But the good news is that we can still avoid the worst consequences if we act quickly.  The AAS and its membership recognize the danger climate change represents to humanity and our world, and to astronomy– as a profession, a hobby, and a cultural good.  We also acknowledge that emissions from astronomy contribute to climate change.  Not everyone is contributing to the problem equally– and those who are the least responsible are suffering the most.  The highest emitters therefore have the greatest ethical obligation to reduce their emissions, which includes astronomy.  **Our profession in general– and the AAS in particular– should work to make it possible for all astronomers to have an equal opportunity to be successful without needing to incur high carbon emissions**, and to preserve astronomy for future generations.

To better understand the problem, a study was completed of the carbon emissions associated with AAS operations and activities over a 12-month period during 2022-23, including the 240th meeting in Pasadena and the 241st meeting in Seattle.  **The analysis found that 84% of total AAS-related emissions are associated with conferences.**  Flights account for 80-90% of this, with the venue, hotels, and catering making up the rest.  The distribution of Sky & Telescope (S&T) magazine is 9% of total AAS emissions.  The balance comes from all other AAS activities, including S&T tours, travel grants, publishing, and facilities.   Professional publications (e.g., AJ Journals and AAS-IOP eBooks) generate less than 1% of total AAS emissions.

We conducted a survey of AAS members to determine their attitudes about climate change. **Respondents overwhelmingly (97%) think that the AAS should reduce its carbon footprint,** with half identifying it as a high priority.  A majority also said the AAS should provide resources on how to effectively teach and communicate about climate change.  Only 25% said that we should purchase carbon offsets.  We also asked questions about the AAS meeting format.  A majority indicated that the following are "very important" to their conference experience: connecting with colleagues and friends, attending talks, meeting with collaborators, and presenting their research.  Our survey results also indicate that there are a range of factors that exclude people from attending in-person meetings, including: concerns about climate change, COVID, lack of funding, work and family responsibilities, and physical impediments to travel and participation.  Online meetings have fewer barriers, and are more diverse and inclusive.  While many expressed frustration with their experiences online, there was also quite a bit of appreciation.  **Respondents identified over thirty virtual conferences that used new technologies and formats that they thought were effective.**  While there is still much to learn, it is clear that online meetings have a bright future.

After much deliberation, our task force created a list of fourteen recommendations, which were then prioritized by a voting process. Each member had five votes to allocate to the fourteen recommendations, which are described in detail in the full report. In order of priority they are:

1. Do not schedule additional in-person meetings before 2030 (18 votes)
2. Innovate the AAS conference model (15 votes)
3. Do not use carbon offsets (7 votes)
4. Move select elements of in-person meetings online (7 votes)
5. Avoid or reduce international meetings or travel (6 votes)
6. Educate AAS members and the broader community on climate change (5 votes)
7. Reduce the emissions associated with in-person meetings (3 votes)
8. Partner with other professional societies (3 votes)
9. Transition to electronic distribution of S&T magazine (1 vote)
10. Support a grant program for testing new interaction technologies (1 vote)
11. Replace AAS in-person work trips with online meetings (1 vote)
12. Divest from fossil fuels and invest in renewable energy (1 vote)
13. Advocate for systemic change to reduce carbon emissions globally (1 vote)
14. Advocate for systemic change to reduce the carbon footprint of astronomy (1 vote)

Our voting is based upon this logic: To make meaningful reductions in its carbon emissions the AAS must either reduce the number of people attending in-person meetings, reduce the frequency of in-person meetings, or reduce (or eliminate) the distance traveled. Analysis indicates that a careful selection of meeting location will only result in at most a 25% reduction. In contrast, **the carbon footprint of a virtual meeting is less than 1% that of in-person**. It is clear that online interaction is the only way to increase participation while meaningfully decreasing emissions. We stress that we are not asking for a return to on-line or hybrid meetings as they were done during the COVID pandemic. Instead, we believe the key to reducing our carbon footprint is to reinvent the traditional conference model– not only because it will reduce emission but also because it will better serve our evolving community.

Each of these recommendations has a different timeline and level of impact. However we urge the AAS leadership to immediately adopt the first recommendation and commit to the second. We recognize that this is a major transition from our traditional working model. But **maintaining the status quo and signing new contracts will tie our hands for the future.** It will make it impossible for us to meet the terms of the Paris Agreement– nor can we pivot to new, more inclusive meeting and interaction modes. Our recommendations are aligned with the Astro2020 Decadal Survey as well as AAS values to disseminate our scientific understanding of the universe, and to do our work in an ethically responsible way. These recommendations also reflect the overwhelming support– from 97% of our membership– to reduce our emissions. Because of their other benefits– particularly in making our society more welcoming to those who traditionally have been excluded– we feel that these are sound decisions, worthy of implementation even if the AAS *wasn't* trying to reduce its carbon footprint. They simply make sense as steps towards a professional society that better serves a broader membership, as our profession evolves to be greener, more inclusive, and more productive.

# Introduction

The American Astronomical Society (AAS) and its membership recognize the danger climate change represents to humanity and our world.  We also recognize the threat that climate change poses to astronomy, as a profession, a hobby, and a cultural good.  And we acknowledge that emissions from professional astronomy– including AAS activities– contribute to climate change.  The [AAS Mission and Vision Statement](#) declares that one of our values is to "accomplish our work using environmentally sensitive actions rooted in scientific understanding."

As part of Strategic Priority 1, the [AAS Strategic Plan for 2021-26](#) identified as a goal to "Develop and implement a plan for the role of the AAS in mitigating climate change."  In particular, it called for the creation of this task force to identify how the AAS can meet the terms of the [Paris Agreement](#).  Adopted in 2015, the Paris Agreement is an international treaty among 195 nations with the goal of avoiding the worst consequences of climate change by keeping the global temperature increase above pre-industrial levels to well below 2°C, and preferably to within 1.5°C.

This task force was established in February 2022 and consists of the following members:

| | | |
|---|---|---|
| Travis Rector (Chair) | Univ. of Alaska Anchorage | tarector@alaska.edu |
| Louis Barbier | NASA | lmbarbier@gmail.com |
| Andrew Couperis | Georgia State Univ. | acouperus1@gsu.edu |
| Rolf Danner | JPL/NASA | rolf.m.danner@jpl.nasa.gov |
| Arika Egan | JHU/APL | arika.egan@jhuapl.edu |
| Paul Green | Smithsonian Astrophys. Obs. | pgreen@cfa.harvard.edu |
| George Jacoby | NOIRLab | george.jacoby@noirlab.edu |
| Jackie Monkiewicz | Arizona St. Univ. | jmonkiew@gmail.com |
| Robert Nikutta | NOIRLab | robert.nikutta@noirlab.edu |
| Karly Pitman | Space Science Inst. | pitman@spacescience.org |
| Michael Rutkowski | Minn. St. Univ., Mankato | michael.rutkowski@mnsu.edu |
| Sarah Tuttle | Univ. of Washington Seattle | tuttlese@uw.edu |
| Anne Virkki | Univ. of Helsinki | annevirk@gmail.com |
| Kat Volk | Planetary Science Institute | kat.volk@gmail.com |

In forming this task force it was recognized that our recommendations needed to be aligned with other goals of the AAS as identified in our Mission and Vision Statement and the Strategic Plan.  Members for the task force were therefore solicited from established AAS working groups and committees.

Since its formation, the task force has held regularly scheduled meetings to discuss our progress.  Meetings have been recorded and notes taken for those unable to attend.  To achieve our goals the AAS contracted with [alectro.io](#) to complete an assessment of the carbon emissions associated with AAS operations and activities.  With the assistance of the American Institute of Physics (AIP) Statistical Research Center we also conducted a survey of AAS members to determine their attitudes towards climate change and potential solutions.  The recommendations shared here are based upon the results of the carbon assessment and member survey.

## Why the Urgency?

It is hard to properly convey the severity of the problem, and the urgency to act. Simply put, climate change is a crisis. It is a different kind of crisis than COVID– of a much larger size and unfolding on longer timescales– but it nonetheless requires our immediate attention. The bad news is that climate change is already causing severe harm. 2023 was the world's warmest year on record, by far (NOAA 2024)[1]. Not only is it hotter, but extreme weather events are more common (e.g., NAS 2016)[2]. Last year the firestorm in Maui, flooding in Chile, extreme blizzards throughout the U.S., and Hurricane Idalia in Florida were all stark reminders of how climate change is affecting us *now*. In 2023 there were 25 weather and climate disaster events with losses exceeding $1 billion each, and that was just in the United States (NOAA 2023)[3].

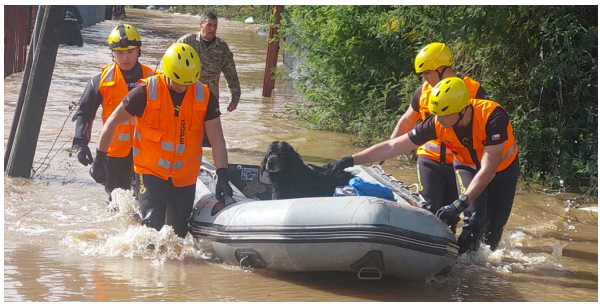
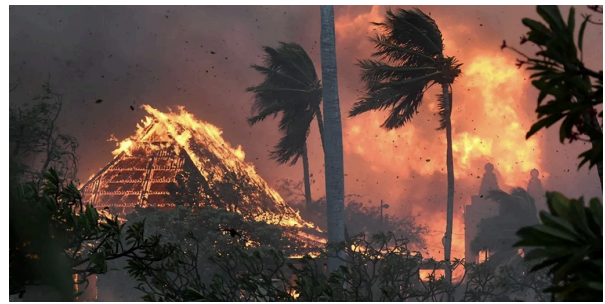
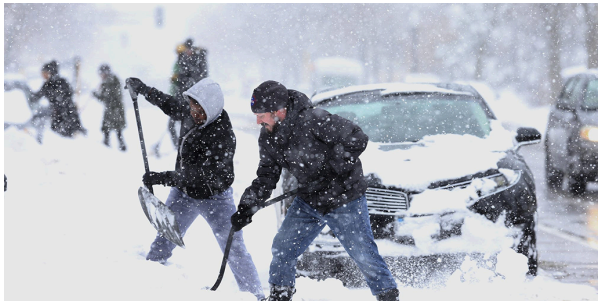
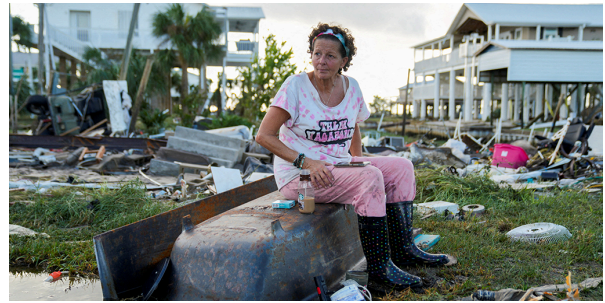

Image sources (upper-left to lower-right): Flooding in Chile, the firestorm in Maui, a blizzard in Minneapolis, and damage from Hurricane Idalia in Florida.

Climate change is also a threat to astronomy. In recent years many of our observatories have been put in danger or damaged by extreme events exacerbated by climate change; e.g., the wildfires at Mt. Stromlo in 2003, Mt. Wilson in 2020[4], and Kitt Peak in 2022[5] caused major damage and disruptions to observing, and threatened the complete loss of the observatories operating on these mountaintops.  Hurricane damage also contributed to the collapse of Arecibo Observatory (Clery 2017)[6].  Changing climate conditions are also affecting the long-term stability of weather patterns (Gómez Toribio et al. 2022)[7], resulting in more lost nights (e.g., van Kooten and Izett 2022[8]; Haslebacher et al. 2022[9]; Seidel et al. 2023[10]), increased sky brightness (Ściężor and Kubala 2014[11]; Steinbring 2022[12]), and degraded image quality (Cantalloube et al. 2020)[13].

---

[1] https://www.noaa.gov/news/2023-was-worlds-warmest-year-on-record-by-far
[2] https://doi.org/10.17226/21852
[3] https://www.ncei.noaa.gov/access/billions/
[4] https://www.chara.gsu.edu/press-release/bobcat-fire-burning-in-the-angeles-national-forest
[5] https://noirlab.edu/public/news/noirlab2213/

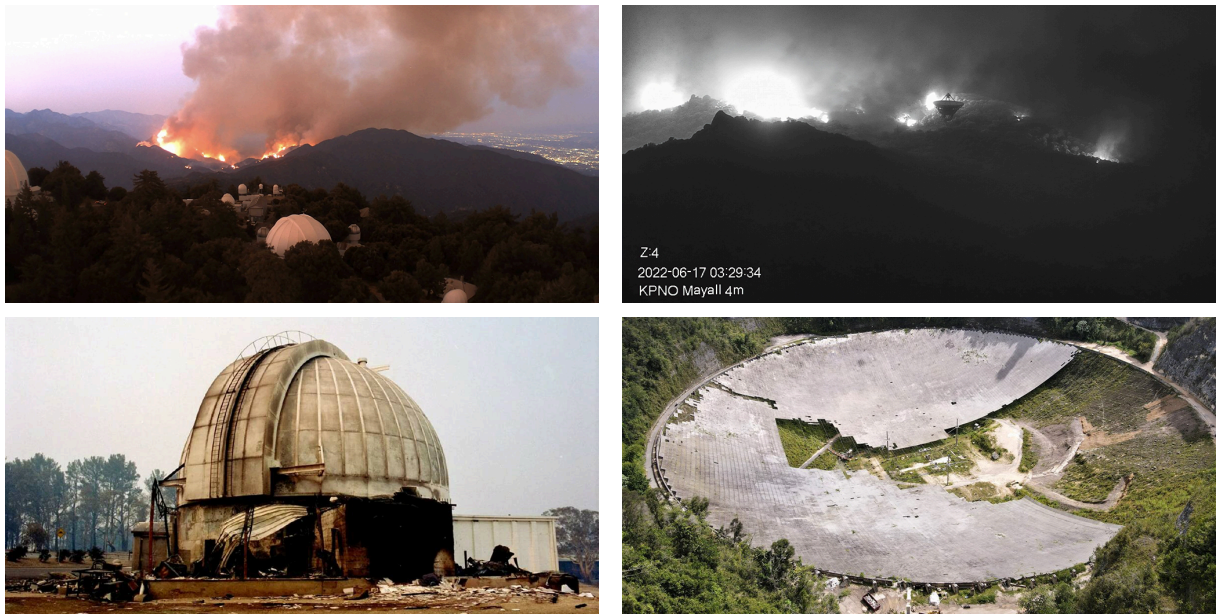

Image sources (upper-left to lower-right): the Bobcat fire on Mt. Wilson, the Contreras fire on Kitt Peak, the remains of Mt. Stromlo Observatory, and the remains of Arecibo Observatory.

The good news is that we can still avoid the worst consequences of climate change, but we have to act fast.  To avoid irreversible "tipping points", climate scientists are calling for us to keep the temperature anomaly at or below 1.5°C.  But to do so, the world needs to rapidly drop its carbon emissions– to 50% of 2019 levels by 2030, with the long-term goal of going carbon neutral by 2050 (Figure 1).  This is the goal of the Paris Agreement, **thus the AAS has set a goal to cut its emissions in half by 2030**.  To learn more about the consequences of exceeding 1.5°C and what is needed to avoid them, please read this IPCC special report.

---

[6] https://doi.org/10.1126/science.aaq0598
[7] https://doi.org/10.48550/arXiv.2103.03917
[8] https://doi.org/10.1088/1538-3873/ac81ec
[9] https://doi.org/10.48550/arXiv.2208.04918
[10] https://doi.org/10.3390/atmos14101511
[11] https://doi.org/10.1093/mnras/stu1577
[12] https://doi.org/10.1088/1538-3873/acac52
[13] https://doi.org/10.1038/s41550-020-1203-3

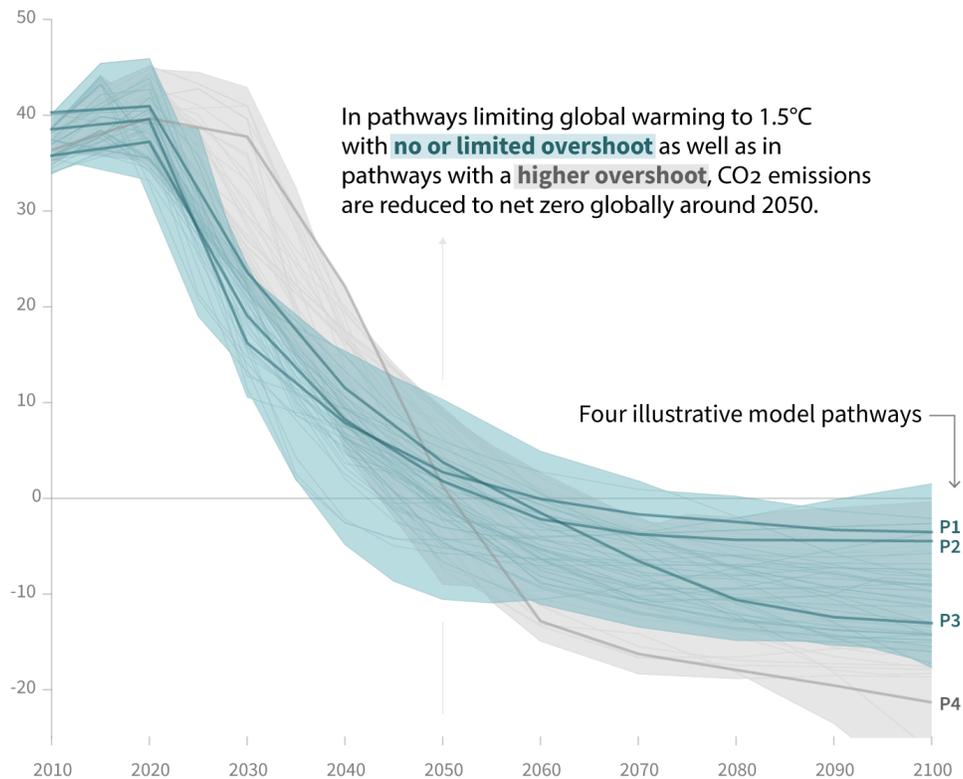

Figure 1: Figure SPM.3a from the IPCC Special Report on 1.5°C. This plot shows the reduction in carbon emissions necessary to avoid the worst consequences of climate change.

When considering solutions, we need to recognize that not everyone is contributing to the problem equally. And that those who are the least responsible are suffering the greatest consequences. The highest emitters therefore have an ethical obligation to reduce their emissions the most. On a per-capita basis, U.S. Americans are among the highest emitters in the world, about three times the global average[14,15]. **And while astronomy is a small profession, on a per-person basis our emissions are more than double the U.S. national average** (Knödlseder et al. 2022)[16]. We are also among the highest emitters in academic fields (e.g., Blanchard et al. 2022)[17]. It is also important to consider that our emissions are not equally distributed amongst all astronomers. In general, senior scientists produce much higher emissions than early-career astronomers (Stevens et al. 2020)[18].

---

[14] https://www.iea.org/commentaries/the-world-s-top-1-of-emitters-produce-over-1000-times-more-co2-than-the-bottom-1#
[15] https://www.statista.com/chart/24306/carbon-emissions-per-capita-by-country/
[16] https://doi.org/10.1038/s41550-022-01612-3
[17] https://doi.org/10.1371/journal.pclm.0000070
[18] https://doi.org/10.1038/s41550-020-1169-1

Astronomers around the world are acknowledging the threat climate change poses, to their field and society at large, and they are taking action[19]. The "big three" sources of emissions in astronomy are from air travel, supercomputing, and the construction and operation of astronomical observatories and facilities. The [Astro2020](#) Pathways to Discovery Decadal Report recommends the following actions to address climate change: " … increase the use of remote observing, hybrid conferences, and remote conferences, to decrease travel impact on carbon emissions and climate change." Many astronomical organizations are working to reduce their emissions on the scale required by the Paris Agreement. For example, the [NOIRLab Environmental Sustainability Program](#) aims to decrease the observatory's annual carbon output by 50% by the end of 2027. ESO also has an [environmental sustainability program](#). And Keck Observatory has set a goal of being [carbon neutral by 2035](#). Additional articles on astronomy's carbon footprint, as well as on efforts to reduce it, can be found in this [Nature Astronomy special issue](#). However the focus of our task force is on AAS-related emissions.

Reducing our emissions will have an impact well beyond our field. Scientists in general are highly regarded (e.g., Pew Research 2022)[20], and of course astronomy is particularly popular. Scientific messages about the need to address climate change are taken more seriously by the public when scientists themselves reduce their emissions (Attari et al. 2016).[21] **We wish to be good stewards of the Earth as well as role models for society at large.**

## Results of Carbon Assessment Study

For the purposes of this task force, alectro.io completed an analysis of the carbon emissions associated with AAS operations and activities over a 12-month period during 2022-23. The analysis includes the 240th AAS meeting in Pasadena and the 241st meeting in Seattle.

As shown in Figure 2, their study finds that **84% of total AAS-related emissions are associated with conferences** and 9% with the distribution of Sky & Telescope (S&T) magazine. The balance comes from all other AAS activities, including S&T tours, travel grants, publishing, and facilities operations. The total AAS emissions are dominated by conferences, and conference-related emissions are dominated by air travel. Flights account for 80-90% of conference emissions, with the venue, hotels, and catering making up the rest.

---

[19] https://www.nytimes.com/2024/05/14/science/astronomy-climate-change.html?smid=url-share
[20] https://www.pewresearch.org/science/2022/10/25/americans-value-u-s-role-as-scientific-leader-but-38-say-country-is-losing-ground-globally/
[21] https://link.springer.com/article/10.1007/s10584-016-1713-2

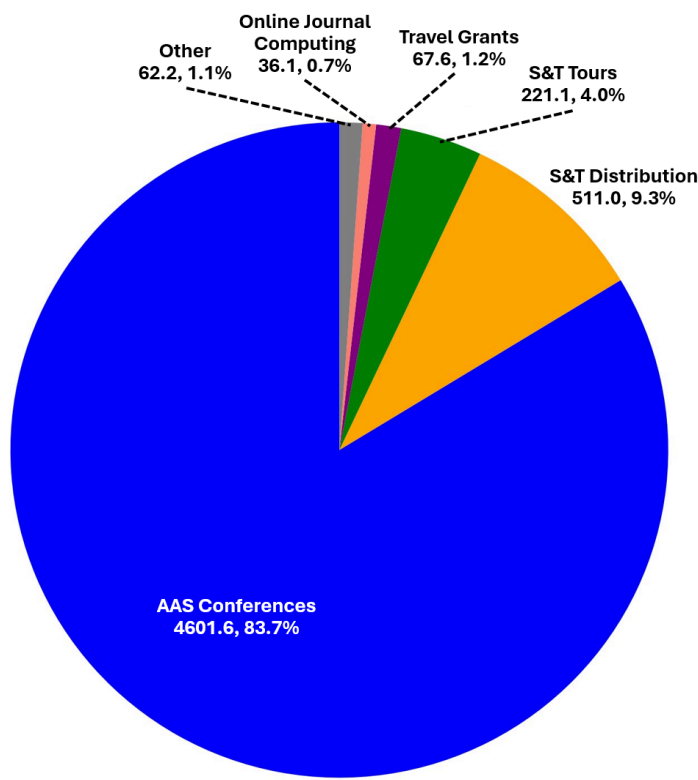

Figure 2: The carbon emissions associated with AAS activities as determined by alectro.io. For each category the annual tons of CO$_2$ emissions equivalent (tCO$_{2e}$) are listed as well as its percentage of total AAS emissions. "AAS conferences" refers to the summer and winter meetings only. "S&T Distribution" is for shipping physical copies of Sky & Telescope magazines. "Travel Grants" refers to the international travel grant program administered by the AAS unrelated to the summer and winter meetings. "Online Journal Computing" is for AAS publications hosted online. "Other" includes AAS facilities, employee commutes, and produced and purchased goods.

The emissions associated with the 241st AAS Meeting in Seattle were determined to be 0.97 tons of CO$_2$ equivalent (tCO$_{2e}$) per in-person attendee. The 240th Meeting in Pasadena had a lower impact of 0.86 tCO$_{2e}$ per in-person attendee. The differences between the two can be attributed primarily to the greater fraction of local attendees at the Pasadena meeting, as there are significantly more astronomers in southern California. These results are consistent with other studies of astronomical meetings (e.g., Burtscher et al. 2020[22]; Réville et al. 2021[23]; Gokus et al. 2024[24]). To put these numbers in a global perspective, **each in-person attendance produces as much emissions as the average annual per-capita carbon footprint in Africa** (Arias et al. 2021)[25].

---

[22] https://doi.org/10.1038/s41550-020-1207-z
[23] https://zenodo.org/record/7467350
[24] https://doi.org/10.1093/pnasnexus/pgae143
[25] https://www.ipcc.ch/report/ar6/wg1/

We note that international travel generates especially high emissions. The international travel grants have a very large footprint of on average 2.6 tCO$_{2e}$ per person. And international trips run by S&T Tours generate 0.8 to 3.0 tCO$_{2e}$ per person depending on the destination.

Of the emissions associated with S&T magazine, the majority is associated not with the actual production of paper copies but with their dissemination. This impact is especially felt with international distribution– while only 7% of readers are outside of the US and Canada, 34% of the impact derives from them. For comparison, professional AAS publications (e.g., AJ Journals and AAS-IOP eBooks) generate only 0.7% of total AAS emissions because they are distributed electronically. **Unsurprisingly, electrons are much easier to move than people or paper.**

# AAS Member Survey Results

The survey was conducted in February 2023. An invitation to participate was sent to half of AAS members in the U.S. (3299 people) and all of the international members (682) for which demographic data were available. 32% and 37% responded respectively, which is typical for AAS surveys. The results are therefore considered to be statistically representative of the AAS membership. Responses were analyzed to look for statistically significant variation as a function of demographic differences, including gender, age, race/ethnicity, level of education, disability, and identity as Lesbian, Gay, Bisexual, Transgender, Queer or Intersex (LGBTQI+). The year of completion of the highest degree was used as a proxy for age.

## Attitudes about Climate Change

AAS members overwhelmingly (98%) expressed concern about climate change's impact on humanity, with 72% saying they are "extremely concerned". These percentages are even higher for more recent graduates. Slightly lower numbers (87%) expressed concern about the impact of climate change on astronomy as a profession, with only 20% saying they are "extremely concerned". These numbers are slightly lower for men but higher for those who have a disability.

**AAS members also overwhelmingly (97%) think that it is important for the AAS to reduce its carbon footprint**, with half identifying it as a high priority. Three-quarters of AAS members think that the AAS should advocate for solutions to climate change. These numbers are higher for more recent graduates, and slightly lower for men.

We asked respondents to indicate which actions AAS could take to help its members address climate change, in order of priority:

|  | Agree |
|---|---|
| Provide resources on ways to reduce one's professional carbon footprint | 71% |
| Provide resources for teaching climate change in informal education settings (e.g., public talks)$^{◇\ †}$ | 59% |
| Provide resources for ways to advocate for climate change solutions◇ † | 58% |
| Provide resources for teaching climate change in formal education settings (e.g., in the classroom)◇ † | 54% |
| Purchase carbon offsets◇ † | 25% |
| AAS should *not* engage in addressing climate change* | 6% |

◇Women and people with other gender identities than men are more likely to agree. (Note: To obtain a statistically meaningful sample, women and people with other gender identities were combined.)
† International members are less likely to agree.
*Only 1% of women and people with other gender identities than men selected this option.

Categories with majority support are labeled in green, whereas those without are shown in red.

## Attitudes about Conferences

Because in-person conference attendance is the primary contributor to the AAS carbon footprint, members were asked about this topic. 71% of the respondents reported attending an AAS meeting in-person in the last five years; and 46% said they had attended virtually during the same time period. 22% said they had not attended in either format.

### Reasons for Attending

Regardless of format, respondents indicated that the following factors were important reasons for participating in AAS meetings, in order of priority:

| | Very Important | Somewhat important | Not important |
|---|---|---|---|
| Connect with colleagues, friends | 76% | 22% | 1% |
| Attend talks | 65% | 33% | 3% |
| Meet with collaborators | 71% | 25% | 4% |
| Present own research | 62% | 33% | 5% |
| Town halls, open house events | 26% | 56% | 18% |
| Participate in workshops◇ | 26% | 55% | 20% |
| Exhibit hall † | 27% | 50% | 24% |
| Career center / job interview* | 31% | 40% | 29% |
| Represent dept/org/project | 29% | 40% | 38% |
| AAS business meetings◇ | 7% | 30% | 63% |

◇More important to women and people with other gender identities than men. (Note: To obtain a statistically meaningful sample, women and people with other gender identities were combined.)

†Less important to international members.
*More important for recent graduates but less important for those with a Ph.D.

The results can be divided into three groups: Respondents indicate that professional and social interaction as well as the dissemination of research (labeled above in green) are the primary reasons for participating in AAS meetings. Other activities (labeled above in yellow) are important but of lower priority. And AAS business meetings are deemed to be of the lowest importance.

Factors Preventing or Hindering Attendance

The survey included questions about factors that might prevent or hinder a respondent from attending a meeting, either in person or virtually. For symmetry, the same questions were asked for both formats. The table below shows what percentage of respondents indicated that a particular factor prevented or hindered attendance.

|  | In-Person | Virtual |
| --- | --- | --- |
| Concerns about COVID-19 | 89% | 27%† |
| High registration costs | 78% | 73% |
| Work responsibilities | 78% | 71% |
| Lack of funding for other expenses | 68% | 33% |
| Concerns about climate change | 52% | 22% |
| Lack of employer support | 38% | 83% |
| Physical barriers to travel and/or participation | 36%* | 24% |
| Other family issues | 33% | 21% |
| Childcare issues | 22% | 14% |

*This value is about 50% for international members and those with a disability.
†Higher levels of concern among international members and recent graduates.

Categories with major (> 30%) differences between virtual and in-person are labeled in red. Lesser (but statistically significant) differences are labeled in yellow. No significant differences are identified in category responses based upon gender or racial differences. We highlight that **52% said that concern about climate change hindered or prevented their attendance at an in-person meeting**. Why is the lack of employer support higher for virtual? Some reported that employers would not cover registration fees for virtual meetings. Others said that employers did not give them time off from other responsibilities while attending virtually. And some indicated that virtual conferences would not be considered for tenure and promotion purposes.

Other factors that hindered or prevented in-person attendance (as given in the open-ended questions) include:
- Not sufficiently beneficial (e.g., don't like meeting format, lacks focus, talks too short, too many parallel sessions, not looking for a job)
- Cost-benefit not worth it (e.g., travel to in-person takes too much time, too expensive)
- Scheduling of meetings interferes with professional obligations (e.g., many reported that the winter meeting often overlaps with the start of spring semester).
- Scheduling of meetings interferes with personal time (e.g., holidays and vacation time)
- Too big (e.g., too many parallel sessions)
- Attending other conferences more important
- Lack of research results to present
- Disagree with masking, vaccination policies
- Visa issues for international participants
- Meeting location (too far away, too expensive, only go to local meetings, etc.)

Other factors that hindered or prevented virtual attendance (from the open-ended questions) include:
- Prior bad experiences with virtual meetings (e.g., Zoom fatigue, hard to stay focused), especially in the context of prolonged online periods during COVID
- Lack of social contact, networking opportunities
- Employer, funding source, etc. wouldn't cover registration costs
- Not recognized for tenure, promotion
- Too expensive
- Having to do normal duties (professional and personal) while attending
- Time zones
- Limited or poor access as a virtual attendee at hybrid meeting (can't access some sessions, low engagement with in-person attendees, technical issues, poor online interfaces, 'impaired' experience)
- Virtual meetings are less "fun"

We also asked respondents to describe health factors that limited or prevented their participation. The list includes:
- Physical limitations that impair/prevent travel to/from and during conferences
- Chronic pain
- Discomfort sitting for long periods of time (esp. in chairs with poor back support)
- Dietary restrictions
- Hearing impairment (esp. in loud rooms)
- Immunosuppression
- Social anxiety
- Neurodivergence
- ADHD

Additional results from the survey are given in Appendix D.

# Meeting Cadences, Locations, and Formats

According to the alectro.io assessment, about 90% of AAS-related emissions are associated with air travel. As discussed in Appendix A, air travel will remain carbon intensive for many decades. And as discussed in Appendix B, carbon offsets are not reliably effective. Thus to make meaningful reductions in its carbon emissions the AAS must either reduce the number of people attending in-person meetings, reduce the frequency of in-person meetings, or reduce (or eliminate) the distance traveled.

Careful consideration needs to be made about the benefits and unintended side effects of different solutions. For example, one option is to simply reduce the number of in-person meetings; e.g., have only one in-person meeting per year, or one every 3 years like the IAU General Assembly. However, winter meetings are usually much larger than summer (about 3000 attendees, compared to 500-1000), with relatively little overlap in attendance. Thus holding a single in-person meeting per year would not reduce emissions alone if everyone simply attends the one meeting held.

A careful choice of host cities has potential to reduce carbon emissions. As discussed in the alectro.io assessment, the 240th Meeting in Pasadena resulted in about 10% less emission per in-person attendee compared to the 241st Meeting in Seattle, as a result of the larger fraction of local participants. A more centralized location could further reduce emissions. Gokus et al. (2024) calculated several scenarios specifically for the AAS. They find that if the 233rd AAS meeting in Seattle had instead been held in a central location (e.g., Chicago or Minneapolis) it would have led to as much as a 25% reduction in emissions.

Another option is the "hub" model. Multiple AAS "regional" meetings– which could be held simultaneously or not– would further reduce carbon emissions. Gokus et al. (2024) calculate that if the 233rd AAS meeting in Seattle had been held in two coastal locations instead (e.g., Los Angeles and Baltimore, with attendees traveling to the closer location), it would have led to as much as a 50% reduction in emissions. Adding hubs in Europe and Asia for international members would have resulted in an additional 5-10% reduction. How a simultaneous multi-hub conference would be structured is unclear, but Parncutt et al. (2021) offer an example. The hub model would fundamentally change the nature of AAS meetings; and would greatly increase the workload for AAS staff. If the meetings were held synchronously, it would introduce problems similar to those encountered with hybrid meetings. For these reasons we don't consider the hub model to be viable, or even desirable.

From a carbon reduction perspective, by far the best option is virtual meetings. While it is true that computers require energy, **the carbon footprint of a virtual meeting is less than 1% that of in-person** (e.g., Burtscher et al. 2020, Tao et al. 2021).

These options are summarized in the table below. Keeping the in-person meeting in the current format but choosing more optimized locations (labeled red) will reduce emissions but not enough to meet the goals of the Paris Agreement. Regional "hub" meetings (yellow) could meet the goals if other measures are taken, including canceling the in-person meetings that are already scheduled. **Only a transition to virtual meetings (green) will make it possible to achieve our goals without canceling meetings for which the AAS is already contractually obligated.**

| Meeting Location | Reduction* |
|---|---|
| Coastal, large astronomer population (e.g., Los Angeles, Washington DC) | ~10% |
| Centrally located, airport hub city (e.g., Chicago, Minneapolis, St. Louis) | ~25% |
| Regional "hub" meetings (participants travel to closest city) | 50-60% |
| Virtual meeting (no air travel) | >99% |

*Reductions are estimated per person relative to the 241st meeting in Seattle.

## Recommendations

As part of the AAS Strategic Plan for 2021-26 the AAS has set the goal of reducing its carbon emissions commensurate with the terms of the Paris Agreement. The Strategic Plan also calls for us to build diverse, equitable, and inclusive (DEI) practices within the astronomical community, and to "improve access to and equitable participation in meetings, events, journals, and all AAS services."

Climate change is fundamentally an inequality issue: The richest 10% are responsible for 50% of emissions, while the poorest 50% cause only 8% of emissions (Khalfan et al. 2023)[26]. Many astronomers are within that top 10%. There is also inequality within our profession. Who has the resources to travel and who doesn't? Who is excluded by the way we work? How can we reduce our carbon footprint *and* make our profession more inclusive? **Our profession in general– and the AAS in particular– should work to make it possible for all astronomers to have an equal opportunity to be successful without needing to incur high carbon emissions.**

Our task force has chosen to adhere to these principles when considering potential solutions. They should:
- Meet the goals of the Paris Agreement.
- Increase participation in AAS activities.
- Enfranchise underrepresented groups.
- Improve opportunities for early-career scientists.
- Be financially and logistically viable for the AAS and its members.

Members of the task force suggested ideas that were ultimately organized into fourteen recommendations. As the far majority of AAS carbon emissions are the result of in-person

---
[26] http://doi.org/10.21201/2023.000001

meetings, our ideas naturally focused there. We then reviewed and prioritized the propositions through a voting process. Each committee member had five votes to allocate to the fourteen recommendations. Voters could choose to allocate more than one vote to their highest priorities. No recommendations were eliminated. The main purpose of this prioritization is to give the AAS leadership a sense of how members of the task force see the priorities. While each voter had different preferences, in general we valued solutions with the highest impact– for the AAS as a whole as well as on a per-person basis. We recommend the following courses of action, *in order of priority*:

These two are by far the top priority (18 and 15 votes each):

## 1. Do not schedule additional in-person meetings before 2030

We want the AAS to recognize the urgency of the situation. As of this writing, the AAS [currently](currently) has in-person winter meetings scheduled through January 2027, with an additional one in 2030. Summer meetings are also already scheduled through June 2025, with one more in 2028. **The only way that the AAS can achieve the terms of the Paris Agreement by 2030 without canceling any existing meetings is if it does not schedule any additional in-person meetings before 2030**. Committing to more in-person meetings will effectively tie our hands for the future. To be clear we are *not* calling for the AAS to return to online and hybrid meetings as they were done during the COVID pandemic. Instead we recommend the AAS use these unscheduled meeting times as an opportunity to experiment with new modes of interaction online.

## 2. Innovate the AAS conference model

We recommend exploring alternative structures and delivery formats for AAS meetings to not only reduce emissions but also to better meet the needs of AAS members.

**In-person meetings are an exclusive affair in the sense that many cannot attend.** Only about a third of AAS members participate in an AAS meeting in a given year. Our survey results find that there are several reasons that hinder or prevent attendance– including cost, work and family obligations, health, and concerns about climate change. Overall these barriers for participation are much higher for in-person meetings. And marginalized groups are excluded more than others. In contrast, virtual conferences can increase participation; e.g., attendance at ASP meetings more than doubled after they transitioned from in-person to online. Virtual meetings also can improve diversity, equity, and inclusion (e.g., Skiles et al. 2021)[27]. The only way that the AAS can achieve its goals of becoming more inclusive as well as reducing its carbon emissions is to increase its use of virtual meetings and events in lieu of in-person.

We recognize that many AAS members are resistant to on-line meetings, as reflected by many of the responses to the open-ended survey questions. This is in large part because of how the format was rushed into use during the COVID pandemic. Without a blueprint on how to successfully host virtual meetings, many organizations– including the AAS– attempted to replicate the in-person experience online.  Considering the rush to transition this is understandable– after all, the AAS only had a few months to transition for the summer 2020 meeting.  We note that while there were many complaints about the AAS virtual meetings, there was also quite a bit of appreciation.  Example quotes from open-ended survey questions:

> "January 2021 was the only virtual AAS I've attended. I gave a talk in a session from my living room, enjoyed the poster session from my couch, but my favorite was logging in to the NSF town hall meeting from the dog park! :)"

> "I attended AAS 237 and 238. I thought they were just generally very good. I've always enjoyed attending AAS meetings in person every few years, but now personal attendance is very difficult for both financial and other reasons, I very much appreciate the possibility of virtual attendance."

---

[27] https://doi.org/10.1038/s41893-021-00823-2

While many comments lament the lack of informal, spontaneous interaction online, several remarked that– with the right structure– such interaction was indeed possible.  Example quotes:

> "Heliophysics2050 is the only virtual workshop I have seen that managed to replicate the informal conversations and discussions common in-person meetings."

> "CoolStars 20.5. The poster hall setup in Gather.town was the most successful virtual replication I've seen for the in-person poster hall equivalent."

With the proper meeting format and moderation, it is possible to create opportunities for people to socially interact, which is commonly cited as one of the most important aspects of in-person meetings.  Such interaction could be organized in many ways– either randomly or based upon certain characteristics (e.g., shared research interests and hobbies, as well as cross-pollinating senior and early-career scientists).

**Online meetings have a bright future.**  The format is new and will continue to evolve as new methodologies and technologies are tried and developed.  A good analogy is that of the early days of television.  Back then people didn't yet know how to effectively create TV shows, so they did what they knew– they did radio shows on TV.  Eventually they learned the power of the new format, as will be the case with online meetings.  Already several formats have been used effectively, and **the survey results list over thirty examples of successful virtual and hybrid meetings**.  Since the pandemic, [The Future of Meetings](#) (TFOM) group has been exploring many different meeting formats and technologies, and we encourage the AAS to collaborate with them.  Specific recommendations on how to improve virtual and hybrid meetings, including example conferences, are given in Appendices C and D.

The following recommendations were identified as important but of lower priority (5-7 votes) than the top two:

### 3. Do not use carbon offsets

As discussed in more detail in Appendix B, we do not endorse the purchase of carbon offsets as we do not trust them to be effective.  Furthermore, only 25% of surveyed AAS members supported their purchase– and many of the survey comments reflect our doubts.  This skepticism is supported by the work of others; e.g., a review of more than 5600 carbon offset projects found that only 2% of them had a high likelihood of resulting in emission reductions that are additional and not over-estimated (Cames et al. 2016)[28].  Despite extensive searching we were unable to find any offset programs that we felt could be reliably trusted to offset emissions.

If carbon offsets are to be used, effective carbon removal is not cheap– costing $100-600 per ton (IPCC 2022)[29].  The cost of necessary carbon offsets per in-person attendee would therefore need to be in this range.  They would need to be included as part of the registration costs for *all* in-person participants, as voluntary participation in offset programs is ineffective (e.g., Berger et al. 2022)[30].  These additional costs would make attendance even more difficult for many, especially as currently many funding agencies do not cover the costs of carbon offsets.  For all of these reasons we do not support their use.

### 4. Move select elements of in-person meetings online

Regardless of the format, it is worthwhile to move elements of meetings online so as to reduce the burden on participants.  AAS meetings– the winter meetings in particular– have become crowded, with a year's worth of activity compressed into 4-5 days.  Below is a list of reasons why people attend AAS conferences, particularly for in-person meetings:
- Present research
- Learn about recent results from others, especially in one-on-one, low-key meet-ups
- Meet with existing collaborators
- Establish new connections/collaborations (within specialty and beyond)
- Socializing and informal networking
- Be part of a professional community, group celebrations
- Identify trends in the field
- Opportunity to interact with observatories, funding agencies
- Exhibit hall (vendors, recruitment, swag, etc.)
- Job search
- Expected by job for retention/promotion
- Professional development events (workshops, "hack-a-thons", etc.)
- Getting away from everyday distractions so you can be immersed (i.e., "in the zone")
- Opportunity to see world / free vacation ("perk of job")
- Prestige (professionally and beyond); e.g., awards / selected talks

---

[28] https://climate.ec.europa.eu/system/files/2017-04/clean_dev_mechanism_en.pdf
[29] https://doi.org/10.1017/9781009157940.004
[30] https://doi.org/10.1016/j.gloenvcha.2022.102470

- Going into communities and interacting with local K-12, public, amateur astro groups
- Identify vendors for products to use at home institutions

**What of these could be done online, and more effectively?** For example, job-hunting activities (e.g., interviews, job fair mixers, and job application workshops) are a major driver for attendance of the winter meeting. However, its timing isn't ideal, as application deadlines for most positions are in the fall. It also is unfair to hold these activities at in-person meetings, as it disadvantages those who are unable to attend. We therefore recommend that these activities be held separately from meetings and done online so that all AAS members can participate equally, and without requiring the associated carbon emissions. Others from the above list could be done virtually as well.

## 5. Avoid or reduce international meetings or travel

International travel typically generates 2-3 times the carbon emissions of trips within North America. AAS should actively and strongly discourage AAS-organized and supported meetings, e.g., the AAS [Topical Conference Series](#) (AASTCS), from being hosted outside North America. This includes S&T tours. It should also refrain from international AAS-related business travel– including staff, board, and committees.

## 6. Educate AAS members and the broader community on climate change and its relation to astronomical knowledge

In alignment with the AAS mission statement to "enhance and share humanity's scientific understanding of the universe," it is important for the AAS to help share the scientific understanding of climate change and its relation to our astronomical knowledge. The science of climate change strongly overlaps with that of astronomy (e.g., the nature of light, greenhouse effect, and Milankovitch cycles). Astronomers have a wide reach. Fraknoi (2001)[31] estimated that every year a quarter million students take an introductory astronomy class in the U.S. at the college level. For many students this is the last formal science they'll receive. Finally, our profession also gives a unique and important perspective– on the rarity of habitable worlds and the vast distances to them. The phrase "There is no Planet B" ultimately is an astronomical one.

While the far majority of astronomers are concerned about climate change, many feel underprepared to teach or discuss it in formal (e.g., classroom) or informal education (e.g., public talk) settings. The AAS can play a key role in addressing this concern by educating our membership on effective methods for teaching and talking about climate change with students and the public. Our survey results indicate the majority of members want the AAS to provide (1) resources for teaching climate change in both formal and informal settings, (2) resources on ways to reduce one's professional carbon footprint, and (3) resources for ways to advocate for climate change solutions. We suggest these efforts could potentially be carried out in collaboration with the [Astronomers for Planet Earth](#) (A4E) group that has already taken strides to collect and create materials within these categories. Resources and information would then be disseminated and advertised to the AAS membership, perhaps through a combination of webpage repositories, email newsletters, virtual workshops, and events at meetings. Such efforts could fall under the umbrella of a new AAS Education Fellow and/or Climate Policy Fellow program, akin to the existing Public Policy Fellows.

---

[31] Fraknoi, A., (2001) "Enrollments in astronomy 101 courses: An update," Astronomy Education Review, 1(1), 121-123.

The task force considers the following recommendations to be important but of the lowest priority (1-3 votes):

### 7. Reduce the emissions associated with in-person meetings

On-line meetings and events are the only means that the AAS can achieve its goals of meaningfully reducing its carbon footprint without reducing participation. However if and when in-person meetings are held these recommendations can help to reduce their emissions:
- Hold meetings in centrally located cities in the mainland U.S. that are easily accessible and minimize travel for the majority of attendees. Distant locations such as Hawai'i, Alaska, and international destinations should be excluded. Cities that are hubs for many airlines (e.g., Chicago and Minneapolis) will further reduce emissions by minimizing connecting flights. This should also be the case for all primarily AAS-supported conferences, e.g., DPS, HEAD, and AASTCS.
- Consider meeting locations with at least some access to rail travel (e.g., along the Northeast Corridor).
- Reduce plastic waste (e.g. badges, lanyards) and single-use items.
- Serve vegetarian/vegan food.
- Choose energy-efficient (i.e., LEED certified) conference venues.
- Reduce or eliminate swag and handout materials.

The American Meteorological Society (AMS) has additional [recommendations](#) worth considering.

### 8. Partner with other professional societies

The AAS should collaborate with other [AIP societies and affiliates](#) on effective strategies to reduce our carbon emissions, as well as for education and advocacy. For example, the AMS has [course packages](#) for teaching climate change at the undergraduate level. AAS staff should have regular meetings with other national (e.g., ASP) and international societies to learn from each other and to coordinate efforts. The AAS should also collaborate with the [TFOM](#) group.

### 9. Advocate for systemic change to reduce carbon emissions globally

While this may be beyond the charge of our task force or the AAS, we note that our survey results find the majority of AAS members (75%)[32] think that the AAS should advocate for public policy related to climate change at the governmental level (e.g., for [carbon fee and dividend](#) legislation).  This governmental advocacy could be similar to existing astronomy and space policy advocacy efforts by the AAS, but with a new focus on climate change policy solutions as well. This could potentially be carried out in part by the creation of a new AAS Climate Policy Fellow alongside the existing Public Policy Fellow program.

---

[32] We note that this number rises to 87% of the membership for non-male-identifying individuals.

The AAS could also investigate options for offering voluntary contributions to reliable climate action groups when registering for meetings, akin to the "Social Offset" options that have been promoted at past AAS meetings.  Note that this does *not* refer to carbon offsets, which we find to generally be unreliable (see Appendix B).  These contributions would instead support climate-specific advocacy groups.  These could take the form of national organizations, such as the [Clean Air Task Force](#), [Project Drawdown](#), and [Natural Resources Defense Council](#), or local organizations near AAS meeting locations that engage in on-the-ground implementations of sustainable climate initiatives.

### 10. Transition to electronic distribution of S&T magazine

The alectro.io assessment indicates that 9% of AAS-related emissions are from the physical distribution of S&T magazine, with a large proportion of that from international distribution. Other options include stopping physical distribution internationally, or perhaps finding ways to print and distribute locally.

### 11. Replace AAS in-person work trips with online meetings

In general the business world has dramatically [reduced its air travel](#) since the COVID pandemic, reflecting a desire to reduce costs and improve the well-being of their employees.  The AAS should strive to reduce AAS business travel as well.  This includes in-person Board of Trustee and committee meetings.  When in-person trips are to be held they should be scheduled less frequently or in conjunction with in-person AAS meetings.

### 12. Support a grant program for testing new interaction technologies

The AAS should consider financially supporting a grant program to purchase or subsidize new software and hardware for participation in online events.  For example, Virtual Reality (VR) headsets have decreased in price (~250 USD as of early 2024 for a Meta Quest 2, which does not require an additional computer to function) and are quite inexpensive in comparison to attending even just one AAS in-person meeting.

### 13. Divest from fossil-fuel-industry supporting investments and invest in renewable-energy industries

According to the 2022 annual report, the AAS has net assets of $13M.  It is likely that at least some of this is invested in portfolios that support fossil-fuel industries (as well as possibly crypto assets that are energy intensive). Opting to actively divest funds from these industries would add to the global economic pressure to move to rising industries that help transitioning the

energy sector to renewable energy sources, while gaining close-to equal or better financial benefit. The power production from renewable sources is already increasing while that from coal is declining in the U.S.[33], a trend that is required for the U.S. to be able to reach its emission goals. However, these goals will not be reached in time without the financial help of companies and organizations. Supporting renewable energy sources in the U.S. will have positive long-term effects in reducing, e.g., emissions of facilities and infrastructure, including supercomputing. Although the reductions specifically in the AAS emissions will be challenging to measure for this action, the impact would extend far beyond the AAS.

## 14. Advocate for systemic change to reduce the carbon footprint of astronomical research

The AAS can advocate for changes to our profession as a whole that enable astronomers and educators to be professionally successful without incurring a large carbon footprint. For example, the AAS could develop an 'Ethics Policy on Climate' to promote best practices in the astronomy community. This policy could include recommendations along the lines of limiting in-person travel to conferences in favor of virtual conferences, promoting the usage of remote and queue observing, and encouraging the use of renewable energy to power observatories.

At the institutional level, the AAS can advocate for systemic changes that help individuals follow these recommendations. For example, the AAS could advocate that astronomical institutions (e.g., universities, observatories, and funding agencies) recognize virtual conferences as legitimate in terms of reimbursement and professional assessment. Ongoing astronomy advocacy efforts by the AAS could also expand to include goals for enabling government grants and funding initiatives to consider carbon emissions and work to reduce them; e.g., fund renewable power installations at observational and research facilities.

# Conclusions

Guided by the results of the alectro.io assessment and the results of our membership survey, our task force has deliberated on how the AAS can meet the terms of the Paris Agreement while meeting the needs of our membership. We have created a list of fourteen recommendations that will help us achieve these goals.

The two top recommendations are to not schedule any additional in-person meetings before 2030 and to use this time to innovate the AAS meeting model. If the AAS continues to sign contracts for meetings it will be impossible to meet the terms of the Paris Agreement, nor can we pivot to new, more inclusive meeting and interaction modes. **In other words, the AAS needs to stop laying tracks until we know where we want to go.** Except for one about education, the middle group of recommendations is also focused on meetings and travel. The third group consists of actions worth doing, but may have less of an immediate impact or are less within the control of the AAS and its members.

---

[33] https://www.eia.gov/outlooks/steo/report/BTL/2023/02-genmix/article.php

The top two recommendations, together with the middle group, form a strong recommendation to change the conference model to create more opportunities to engage online, and to make in-person meetings less frequent and require less travel. We feel that these are sound recommendations, worthy of implementation even if the AAS *wasn't* trying to reduce our carbon footprint. They simply make sense as steps towards meetings that serve a broader membership better, as we reinvent meetings for the future.

These recommendations reflect our understanding of the problem of climate change and the urgency needed. They are aligned with the AAS values to disseminate our scientific understanding of the universe, and to do our work in an inclusive, ethically responsible way. These recommendations also reflect the overwhelming support– from 97% of our membership– to reduce our emissions.

We note this is the first step in a long-term process– for the AAS and for the world– to eventually be carbon neutral. These recommendations are based on a snapshot in time within a rapidly changing global landscape– we therefore recommend a re-assessment of AAS climate action around 2030.

# Appendices

## Appendix A: The Future of Air Travel

Considering the large emissions produced by air travel, we investigated whether or not the airline industry will be able to decarbonize on the relevant timescales. The aviation sector is currently responsible for about 3% of total $CO_2$ emissions (Statista 2022)[34] and about 5% of total radiative forcing due to non-$CO_2$ impacts such as contrails at high altitude (Lee et al. 2021)[35]. Compared to other transportation sectors aviation's footprint is relatively small, but on a per person basis it is enormous. Air travel is an activity of the privileged few. Globally, 1% of people caused half of aviation emissions in 2018, while 89% did not fly at all (Gössling and Humpe 2020)[36]. Even in developed countries, it is an elite minority of frequent flyers that cause most of the carbon emissions from aviation. For example, in the U.S., 12% of people took 66% of all flights (Hopkinson and Cairns 2020)[37]. For some people– including many astronomers– air travel is the largest portion of their carbon footprint.

Unfortunately the aviation sector will be one of the hardest to decarbonize. Because its primary expense is fuel, the aviation industry has always been keen on reducing usage. The average fuel used by new aircraft on a per passenger-km basis fell approximately 45% from 1968 to 2014, with a long-term trend of 1.1% improvement per year that is expected to continue (Kharina and Rutherford 2015)[38]. However this trend is far too slow to meet the goals of the Paris Agreement.

---

[34] https://www.statista.com/statistics/655057/fuel-consumption-of-airlines-worldwide/
[35] https://doi.org/10.1016/j.atmosenv.2020.117834
[36] https://doi.org/10.1016/j.gloenvcha.2020.102194
[37] Hopkinson L and Cairns S (2020) Elite Status: global inequalities in flying. Report for Possible, March 2021. https://www.wearepossible.org/latest-news/elite-status-how-a-small-minority-around-the-world-take-an-unfair-share-of-flights

One option being pursued is to replace traditional jet fuel with sustainable aviation fuels (SAFs), which are generated from renewable hydrocarbon sources. The production of SAFs does require energy, but– depending on the type of SAF– carbon emissions are reduced by up to 80% (de Jong et al. 2017)[39]. SAFs are called "drop-in fuels" because they have nearly the same chemical and physical characteristics of conventional jet fuel, meaning they can be used in existing aircraft and airport infrastructure. This is particularly important as airplanes can have operational lifetimes of up to 30 years, meaning that SAFs are the only way to significantly decrease carbon emissions without decommissioning aircraft prematurely. However SAFs made up only about 0.1% of total fuel used in 2019. Some airlines have made non-binding pledges to use SAF for up to 10% of their operations by 2030. Cost of production is a major limiting factor. As of early 2020, SAFs cost more than twice that of traditional jet fuel[40]. Large scale deployment of SAFs is a serious challenge, as it requires large investments in new production facilities, strong reduction in production costs, and considerable investments in certification for usage in aircraft (Chiaramonti 2019)[41].

Another option is to develop electric or hydrogen-powered aircraft. Because of its high specific energy and relative ease of storage, petroleum will be very difficult to replace. The specific energy of batteries is the major constraint for battery-powered aviation rather than cost (Viswanathan et al. 2022)[42]. Petroleum has a specific energy about fifty times higher than lithium-ion batteries. Furthermore, battery weight doesn't "burn off" during a flight as does jet fuel. Based upon current battery technologies, electric aircraft will be limited to distances under 500-1000 km, however about 95% of $CO_2$ emissions are from aircraft that fly longer distances (World Economic Forum 2020)[43]. Hydrogen fuel cells could also be used to power electric aircraft, but face similar range and operational issues. Hydrogen combustion is an option that may be feasible for medium to long-haul flights (i.e., under 10,000 km). While hydrogen combustion does not release $CO_2$, it produces water vapor that, at altitude, would result in additional radiative forcing. It is estimated that a hydrogen aircraft therefore could have a 50-75% reduction over current aircraft. Conversion of existing aircraft designs to hydrogen are underway, with certification as early as 2025[44]. Aircraft manufacturer Airbus has announced plans to develop hydrogen-powered commercial aircraft to be in operation by 2035[45].

For these reasons SAF, electric, and hydrogen-powered aircraft are not expected to play a major role in carbon emissions over the next decade or longer.

---

[38] https://trid.trb.org/view/1372314
[39] https://doi.org/10.1186/s13068-017-0739-7
[40] https://www.flyingmag.com/could-saf-be-a-cost-effective-solution-to-rising-aviation-fuel-prices/
[41] https://doi.org/10.1016/j.egypro.2019.01.308
[42] https://doi.org/10.1038/s41586-021-04139-1
[43] https://www3.weforum.org/docs/WEF_Clean_Skies_Tomorrow_SAF_Analytics_2020.pdf
[44] https://electrek.co/2023/03/02/universal-hydrogen-passenger-hydrogen-electric-plane-maiden-flight/
[45] https://www.airbus.com/en/innovation/zero-emission/hydrogen/zeroe

## Appendix B: Carbon Offsets

A good question is whether or not the AAS or its members should purchase carbon offsets to neutralize the emissions incurred by attending in-person meetings.  Carbon offsets are intended to fund projects that either lower greenhouse emissions or sequester them from the atmosphere.  Such projects include forestry (e.g., reforestation, or avoided conversion of existing forested lands), $CO_2$ or methane capture (e.g., from landfills or livestock), construction of renewable energy sources, or investment in more efficient appliances  (e.g., improved cookstoves).  The appropriate pricing of carbon offsets is unclear and varies considerably based upon assumptions.  While carbon offsets are often sold at prices much lower than this, estimates of the true cost to prevent or sequester greenhouse gas (GHG) emissions range from $100-600 ton$^{-1}$ (e.g., IPCC 2022)[46].

There is considerable skepticism about the usefulness of offsets as a GHG reduction tool.  If done properly, they can be a means to reduce emissions.  But if not they can actually make the situation worse– as people will think the problem is being addressed when it is not.  In the worst-case scenario, carbon offsets are merely used to assuage guilt without having much benefit.  To be effective, carbon offsets must absorb or prevent emissions equivalent to the activity in question.  The purchased offset needs to result in a reduction that would otherwise not have occurred; e.g., it does no good to protect a tract of forest land if it wasn't going to be harvested in the first place– or if another tract is harvested instead.  The offset should also be durable and effective in the short term; e.g., it can take a planted tree decades to grow to a size where it effectively sequesters $CO_2$.   And trees are vulnerable to disease, drought, and fire.  Finally, the effectiveness of the offset needs to be verified; e.g., a ProPublica report found that loggers cut down trees in Brazilian forests even though offsets were sold to protect them (Song 2019).[47]  Likewise, Calel et al. (2021)[48] found that carbon offsets were often allocated to projects that would very likely have been built without assistance (i.e., these offsets did not result in "additional" reductions).  A review of more than 5600 carbon offset projects found that only 2% of the projects had a high likelihood of resulting in emission reductions that are additional and not over-estimated (Cames et al. 2016)[49].  Our task force tried to identify carbon offsets that we felt confident would be effective but failed to do so– much like the efforts of others (e.g., Dillon 2023)[50].  For these reasons we do not support the purchase of carbon offsets.

One idea the task force investigated was for the AAS to create its own carbon offset program.  It could collect "green fees" as part of registration and invest directly in carbon reduction projects.  The funds could be managed like Education & Professional Development (EPD) mini-grants, and AAS members could apply for projects; e.g., to install solar panels at an observatory or institute.  Funded projects would then be expected to verify the reduction in emissions.  But such a program would be expensive.  Not including the overhead costs associated with managing such a program, by our calculations it would cost $100-400 ton$^{-1}$ offset– consistent with IPCC estimates.  The amount of the green fee collected from attendees could be adjusted by the AAS to achieve the desired offset level.  Beyond direct offsetting, the revenue could also be used for other sustainability efforts, such as reducing waste at meetings, programs to teach astronomers how to effectively communicate about climate change in their classes, or subsidizing virtual meetings.  However– like other carbon offset programs– the $CO_2$ reduction impact of such a program would be hard to measure.  For these reasons we do not recommend that the AAS create its own carbon offset program.

---

[46] https://doi.org/10.1017/9781009157940.004
[47] https://features.propublica.org/brazil-carbon-offsets/inconvenient-truth-carbon-credits-dont-work-deforestation-redd-acre-cambodia/
[48] http://eprints.lse.ac.uk/112803/1/GRI_do_carbon_offsets_offset_carbon_paper_371.pdf
[49] https://climate.ec.europa.eu/system/files/2017-04/clean_dev_mechanism_en.pdf
[50] https://www.nytimes.com/wirecutter/reviews/buying-carbon-offsets-for-your-flight-doesnt-help/

*"Actual carbon offset plans rarely seem to be effective ways of sequestering or reducing atmospheric carbon and in my opinion should be viewed with skepticism."* - Anonymous AAS Member

## Appendix C: Recommendations for conference formats

### Recommendations for Virtual Meetings

Below is a list of "best practice" ideas for virtual meetings and events:
- Use online conference platforms (e.g., Whova) to organize meetings.
- Use interaction tools (e.g., Gather.town, Slido) for social interaction and poster sessions.
- Use productivity tools (e.g., Slack) to interact with speakers, other participants before, during, or after sessions.
- Pre-recorded talks.  These allow participants to watch (and re-watch) content asynchronously (also rewind and faster playback speed).  Synchronous elements of the conference can then focus on interactive elements.
- Use session moderators to encourage community participation and keep engagement levels high in the virtual environment.
- Include interactive elements (e.g., polls, Q&A, "idea boards")
- Greater focus on discussion (e.g., longer Q&A sessions, panel discussions)
- Zoom breakout rooms for small-group discussions
- Ask questions over Slack (or other platform) that allows questions to be "upvoted" so that most interesting questions are addressed first.  Also gives the presenter time to think out answers.
- Shorter sessions (2-4 hours a day) to accommodate other aspects of life
- Time-staggered sessions (e.g., morning and evening sessions) to deal with time zones
- Smaller, topical meetings.
- Use non-traditional formats such as "unconferences" that emphasize facilitating informal/networking interactions (Budd et al. 2015)[51].
- Use contests and prizes to encourage participation.
- "View random poster/talk" button can create spontaneous opportunities to meet/interact.
- Good tech support to keep things running smoothly.

---

[51] https://doi.org/10.1371/journal.pcbi.1003905

- Maximize opportunities for early career scientists to collaborate, connect, and advance into leadership positions.

Additional suggestions from survey responses are included in Appendix D.

### Recommendations for Hybrid Meetings

The hybrid format is challenging for several reasons:
- Online participants may not be able to participate equally to those in person.
- The ideal format for an in-person meeting is not the same as for virtual.
- A/V equipment costs in conference centers are ludicrous.
- Effectively doubles workload on AAS staff.
- Members with shallower financial pockets may be forever tied to the virtual component only, while more wealthy members can always participate in person.

If hybrid is to be pursued:
- Invest in high-quality audio and visual.
- Ensure virtual participants are fully included (e.g., prioritize them for Q&A).
- Improve interaction between in-person and virtual participants.
- Find outside IT support vendors that are less expensive

## Appendix D:  Additional survey information

Below are additional results from our survey of AAS members:

### Hybrid and Virtual Meeting Format

Survey participants were asked about their thoughts regarding their experience with the AAS "hybrid" meeting format, where people can choose to attend in person or virtually.  Seventy-five respondents said they had virtually attended the Summer 2022 AAS hybrid meeting. Of these, 27% indicated they could participate fully, and 4% indicated they could not.  The majority (69%) said "it depended." When asked whether they felt included, 22% indicated they felt mostly included, and 25% told they felt mostly excluded.  About half said "it depended."

Participants were also asked about their interest in different potential changes to *virtual meetings*.  The table below lists them in order of support:

|  | Support | Neutral | Don't Support |
|---|---|---|---|
| Reduced registration costs*◇ | 80% | 17% | 3% |
| Recorded live sessions for later viewing* | 80% | 14% | 6% |
| Childcare stipends*◇ | 58% | 32% | 10% |
| Add social networking opportunities* | 54% | 37% | 9% |
| Half-day meetings* | 53% | 30% | 17% |
| Pre-recorded talks with live Q&A sessions | 44% | 32% | 24% |
| Shorten meeting to 2-3 days* | 35% | 37% | 28% |
| Meetings with shorter sessions*◇ | 32% | 40% | 27% |
| Spread meeting activities over longer period | 26% | 22% | 52% |
| More than two meetings per year | 12% | 25% | 63% |
| Meetings with longer sessions | 10% | 37% | 53% |

*More support from women and people with other gender identities when compared to men. (Note: To obtain a statistically meaningful sample, women and people with other gender identities were combined.)
◇More support from recent graduates.

Again the results can be divided into three groups: Those labeled in green were supported by more than half of respondents. Those in yellow received more support than not, but a substantial fraction were neutral. And those in red were not supported by the majority.

### Successful Virtual and Hybrid Meetings

We asked respondents to share ideas on how to improve virtual and hybrid meetings, as well as give examples of other (non-AAS) virtual and hybrid conferences they had attended that they felt were successful. Over thirty meetings were mentioned. These include:
- DPS and DDA meetings (2020–2022)
- TESS Science Conference 2 (2021)
- ADASS
- SAZERAC meetings
- Cosmology from Home
- Five Years after HL Tau
- Fundamentals of Gaseous Halos (8-week workshop)
- IAU General Assembly 2022
- Habitable Worlds 2021
- Preventing Harassment in Science 2020
- Galaxy Cluster Formation II
- Sharpest Eyes on the Sky
- Titan Through Time
- NICER Data Analysis Workshop 2021
- Iid2022
- 40 Years of the VLA
- Astronomers Turned Data Scientists
- Exoplanets III
- Cool Stars 20.5
- Streams21 - Constraints on Dark Matter
- DotAstronomy
- LPSC 2023
- ASP 2021

- [Virgo Meeting 2021](#)
- [ALAN 2021](#)
- [Planet-forming Disks: From Surveys to Answers](#)
- [CAP 2022](#)
- [Heliophysics 2050](#)
- [Edinburgh Women in Space Conference 2021](#)
- [PoSTER](#)
- [Gemini Science Meeting 2021](#)
- [A4E 2022 Symposium](#)

### Meeting Cadence

We asked respondents how often AAS meetings should be held in person. While a majority wish to keep the winter meetings yearly, there is greater support for holding summer meetings less often:

|                 | Winter | Summer |
|-----------------|--------|--------|
| Once a year     | 56%    | 30%    |
| Every 2-3 years | 22%    | 36%    |
| Never           | 3%     | 12%    |
| Don't know      | 19%    | 22%    |

In hindsight these results are hard to interpret because it depends on what assumptions are made. For example, it would be unfair to those on the job market to hold meetings less frequently than once a year if employment-related activities (e.g., dissertation talks, job interviews) are held only at meetings.

### Open Ended Comments

Finally, open-ended questions were included to give survey participants an opportunity to share their thoughts. We include recurring and relevant ones here.

#### Open-ended comments regarding virtual, hybrid, and in-person meetings

We also asked what they liked about these meetings and for recommendations on how to improve the virtual experience. Responses include:
- Create social networking opportunities, structured and spontaneous
- Access to all talks, not just plenaries
- More interactive sessions
- Have sessions outside of normal times so participants can work around their home and work schedules, as well as to accommodate different time zones.

- Better enabling and encouraging interaction from virtual participants, such as making it easier to interact in real time with poster presenters, speakers during/after a session
- Restructuring talk session; e.g., after each session have a virtual breakout room to chat with others interested in the topic.
- Record all talks so they can be viewed at a later day for all registered users
- More asynchronous elements
- Train presenters on how to prepare and deliver effective presentations
- Use a virtual social environment like [gather.town](gather.town) to facilitate social interaction.
- Shorter days spread over longer periods of time
- Try talks in other formats (e.g., audio-only podcasts, "lightning talks").
- Better use of Slack; e.g., have channels devoted to particular topics, sessions
- Chat rooms on [Discord](Discord) specific to conference sessions
- Use a "3D" interactive platform (e.g., virtual reality environment such as [Glue](Glue))
- Provide virtual whiteboards for discussion sessions or Q & A sessions.
- Online interfaces that enable quick one-on-one virtual connections (e.g., gather.town)
- Full accessibility support, including closed captioning
- Better sound quality
- Better IT support

Open-ended comments regarding AAS hybrid conference experience include:
- Hybrid meetings create an uneven playing field. "The whole meeting was very slanted to in-person participation. It's almost as if I never attended."
- More difficult to fully participate in. "I communicated via Slack and didn't get proper answers."
- Plenary lecture participation was as good as in person. Other events like posters, and contributed talks, were extremely hard to follow virtually.
- Similar to fully virtual conferences, remote attendees at the hybrid conference still struggled with normal work/personal obligations while trying to attend, time zones, and lacking social / networking / colleague interaction.

Open-ended comments not regarding AAS in-person conferences include:

- Reserve front row seats for hard-of-hearing individuals.
- Maintain designated "quiet rooms" for sensory-sensitive individuals (or anyone) to use during in person meetings, and make these widely advertised and clearly indicated.
- Create designated "community rooms" to enable community-building within subgroups, such as LGBT+ spaces, non-university (high school / grade school / planetarium) educator spaces, neurodivergent spaces, grad student spaces, etc.

Other open-ended comments and concerns from survey

While not representative of respondent attitudes overall, these comments occurred frequently enough to warrant further discussion:

**The AAS should "stay in its lane" and focus on astronomy-specific issues only, and we are taking a risk straying into the "political sphere" of climate change.**

*"AAS should exist to focus on astronomy activities, not climate change."* - Anonymous AAS Member

Not all support it, but a majority of our survey respondents agreed the AAS should take action to address climate change from a variety of educational angles, with 97% indicating it is important for the AAS to reduce its carbon footprint and 75% agreeing that the AAS should advocate for public policy related to climate change. The arena of climate change is indeed political, but it's important to note that it was not politicized by scientists but by interest groups who stand to lose financially by the reduced use of fossil fuels. We also note that the majority of Americans are concerned or alarmed about the threat of climate change (as of 2022)[52]. Furthermore, the AAS has engaged in other politically charged topics (e.g., LGBTQI+ rights) that affect the wellbeing and productivity of AAS members.

**As astronomers, we do not know climate science well enough to teach it.**

*"AAS members are not environmental science experts per se, so I don't think AAS should be trying to educate the public, industry, or politicians about climate change in general."* - Anonymous AAS Member

While we are not all experts in all areas of climate science, the science of climate change greatly overlaps that of astronomy. Many if not most astronomers have formal training in at least some of the areas of radiative transfer, planetary science, atomic physics, exoplanetary science, thermal physics, and the like, all of which can connect to discussions of climate change in various ways. Education and outreach about climate change can also take many forms, and at its most basic level requires minimal technical explanation if communicated well. If some still feel lacking, this is an area where training or materials could be offered through the AAS or affiliated professional organizations. In addition, regardless of our education level in climate science directly, astronomers are still scientists, and using our voice can both demonstrate solidarity with our climate science colleagues while also helping to dispel the notion that some may be "in it for the money". Finally, astronomers hold a strong interest and trust from the public that few other fields can claim, putting us in a unique position to communicate more effectively with the public about climate related topics than many other scientists.

*"Astronomers get respect in the community as scientists concerned about the sky and the atmosphere. Even though we may not be climate scientists per se, we should be leaders in combating climate change both within and without our profession. The AAS should make it as easy as possible for astronomers to communicate both the dangers and plausible mitigations of climate change."* - Anonymous AAS Member

---

[52] https://climatecommunication.yale.edu/about/projects/global-warmings-six-americas/

**Other organizations more closely affiliated with climate science are already advocating for change, so the AAS doesn't need to.**

> "*Climate change is well covered by other professional organizations in whose domain such advocacy properly lives.*" - Anonymous AAS Member

Many other organizations, including some affiliated with the AAS, are outspoken voices for addressing climate change. How much quicker might their message be heeded if we were to shout with them, perhaps in coordinated fashion? How severe does climate change need to be before it warrants one more organization joining the call to action? Or a second? Or a third? Rather than relegating the responsibility to others, this instead presents an opportunity to collaborate with our sibling organizations for more cohesive and improved efforts.

> "*AAS should coordinate with other professional societies. Astronomy is not unique in how it should react to climate change.*" - Anonymous AAS Member

**The carbon footprints of the AAS and astronomy in general are too small to be important on the global scale. Why change meetings when air travel is a small portion of global emissions and the flights might happen anyways?**

> "*While I am extremely concerned about climate change, the solutions lie beyond AAS. AAS can set an example and advocate generally for policies that foster climate friendly technologies, but the carbon footprint of the astronomy community is tiny compared to the global footprint.*" - Anonymous AAS Member

It's true that on a global scale astronomy has a very small carbon footprint, and if astronomy reduces its footprint *that alone* would not significantly affect global emissions. Everyone must do their part, and in particular those with a larger footprint have a greater responsibility. While astronomy is a small profession, on a per-person basis our footprint is quite large. The enhanced emissions of astronomers is primarily due to our travel, supercomputing, and observational infrastructure. For perspective, the U.S. emitted about 15 $tCO_2$ per capita in 2021 (Our World in Data 2023)[53], which is among the highest rates in the world– about twice that of Europe and more than three times the global average. Rough estimates by Marshall et al. (2009)[54] found that US astronomers consume about 50% more energy per person than the average American. Astronomers at the Max Planck Institute for Astronomy emitted an average of 18.1 $tCO_{2e}$ per researcher in 2018 (Jahnke et al. 2020)[55]. And in recent years Australian astronomers emitted an estimated ≥37 $tCO_{2e}$ per astronomer per year (Stevens et al. 2020)[56]. The cumulative footprint of ground- and space-based astronomy infrastructure is reported by Knödlseder et al. (2022)[57] to be 36.6 ± 14.0 $tCO_{2e}$ per year per astronomer.

---

[53] https://ourworldindata.org/grapher/co-emissions-per-capita?tab=chart&country=~USA
[54] https://doi.org/10.48550/arXiv.0903.3384
[55] https://doi.org/10.1038/s41550-020-1202-4
[56] https://doi.org/10.1038/s41550-020-1169-1
[57] https://doi.org/10.1038/s41550-022-01612-3

Astronomy also has a large footprint compared to other natural sciences. Data from a large survey of French researchers also found that astronomers on average had the largest plane travel footprint of all the many scientific disciplines considered (Blanchard et al. 2022)[58] — for example, the average astronomer surveyed had ~55-60% more plane travel by distance in 2019 compared to physicists, and roughly 2.5x more than chemists.

Lastly, regardless of the direct $CO_2$ impact, it is still worth changing to align our professional actions to be in accordance with the severity of the climate crisis and evidence at hand. The public, other professional societies, and people around each of us see what we do — we have an opportunity and a responsibility as privileged individuals to send a clear signal to all that helps raise awareness and motivate action in their own arenas. This in turn further propagates, gradually driving the societal and political system-wide changes we need. It's not individual versus systemic action, it's both helping drive each other. Many emphasize the importance of voting in democratic processes, but by analogy, voting despite knowing you likely won't be the deciding ballot is in many ways similar to reducing astronomy's $CO_2$ footprint despite knowing we won't single-handedly overturn global emissions.

*"If scientists do not change behavior to reduce climate impacts, why would we expect the general public to believe in a climate emergency?"* – Anonymous AAS Member

**If we transition to virtual meetings this could have a negative impact on astronomy.**

*"I recognize that there's a little bit of inconsistency, but I really think that astronomers do their job better when there's an opportunity to meet in person. Sure there's carbon generated by flying to a meeting, but our field will be much poorer if we don't meet in the same room. I would definitely worry that keeping too many meetings in virtual format will turn off people to our field, especially those at the non-elite institutions and those from traditionally-underrepresented demographic groups."* – Anonymous AAS Member

Many of the open-ended comments expressed dissatisfaction with virtual conferences, particularly that they can provide as good of opportunities for interaction and networking as in-person meetings. However many also expressed appreciation for the online format, and optimism that it can be more effective. Already there are many examples of highly successful online conferences, which are included in a separate section of this report. And the online format will continue to improve. Overall, we think that with proper experimentation and refinement virtual gatherings will be able to offer an increase in inclusion and accessibility while giving a suitable avenue for research dissemination and interpersonal collaboration.

*"The AAS should build on the innovations that happened during the pandemic shutdown in enabling virtual conferences. Not only do virtual conferences and symposia reduce or eliminate carbon impact (and also eliminate wasteful spending), they vastly improve access for marginalized researchers."* – Anonymous AAS Member

---

[58] https://doi.org/10.1371/journal.pclm.0000070

**Fears that traveling less will hurt one's career**

"There also needs to be some soul searching about traveling so often. A lot of astronomers are on planes 2x a month, and are afraid to cut back for fear that it will hurt their careers." — Anonymous AAS Member

Historically astronomy has been a profession that requires a lot of travel– to observatories, conferences, review panels, and invited talks.  In many ways one's frequent flier status served as a proxy for one's success.  It's a valid concern that reducing travel will impact one's ability to advance.  It is therefore incumbent upon us to make it possible for people to be successful in astronomy without incurring a large carbon footprint.  Already this is being done with techniques such as remote, queue, and robotic observing as well as remote review panels.  And online meetings are also providing opportunities for people to network, particularly for those who might not have been able to attend in person.  The challenge is for us to improve the online format so that it is effectively meeting the needs of astronomers, particularly those early in their careers.

"I also met a few people online (in slack chat rooms) with related interests from other countries, including a new collaborator, who would not have been able to attend an in-person meeting due to prohibitive travel expenses."  – Anonymous AAS Member

## Selected Quotes from Survey Results

Below is a collection of selected quotes that reflect different ideas, attitudes, and concerns as expressed in the open-ended comments of the survey.  Like all of the open-ended comments, they should not be considered to be statistically representative.  The comments are divided by topic but are otherwise unsorted.

### Climate Advocacy

"I'd love to see the society that represents me have the courage to say we understand the science, and we stand behind measures to avoid utter disaster."

"During the previous two world wars, members of the scientific community were actively transitioned out of performing their regular research and into projects that aided those war efforts. Our view of the efforts to mitigate the effects of climate change should be similar and we should, as a community, be transitioning some (and more) of our research time into research on decarbonization efforts."

"I am an astronomer, not a climatologist – I have no business claiming expertise outside of my field."

"I think it would be much more appropriate to have some cutting edge climate change talks at the AAS meetings than build up an expectation that astronomers know enough climate science to teach it."

"This is an "all scientists on deck" kind of moment. I don't think any of us can just stand by at this point."

"The Earth is a planet. The greenhouse effect is a fundamental tool for assessing a planetary atmosphere. As astronomers we know these things well. We therefore have a responsibility to act responsibly ourselves and to engage in public education."

"We tell ourselves that *our* industry is important enough to justify the travel, as do all other industries."

"As an organization, AAS has more power to affect climate change than individuals."

"When someone finds out that I am an astronomer, they often want to know if I think there is life beyond Earth. The answer is simple: I don't know. But I do know we are destroying the only life in the Universe we are certain of."

"My home burned up in the Beachie Creek wildfire. I've spent the last two years trying to recover from the loss of all of my personal property. And yes, the wildfire was the result at some level of climate change."

"I think that education is by far the best way of changing attitudes to climate change."

"AAS should coordinate with other professional societies."

"It would be very interesting if the AAS could advocate to funding agencies for the ability to itemize and include things like carbon offsets in grant reimbursements."

"The AAS should be playing a far bigger role in training individuals to communicate about climate change in their teaching and supporting the inclusion of climate change topics in all introductory astronomy courses."

"The AAS should advocate for more environment-friendly methods of construction and operation for Astronomy facilities, including but not limited to ground-based observatories. The AAS should help the community persuade funding agencies to invest in renewable energy solutions like solar panels to fully/partly power these facilities."

"How do we set an example rather than just preaching to others?"

Virtual, Hybrid, and In-Person Conferences

"I have been attending the GBO virtual colloquia and also the VLA virtual colloquia (when they were still being offered) and it has been very valuable. I have a hearing disability, and unlike in person, I can hear the speaker during the virtual talk and, thanks to the Chat, know what each question is during the Q&A (as opposed to not being able to hear the majority of the questions). I can also ask questions without feeling embarrassed or intimidated, since the speaker knows nothing about who is asking the question (rather like blind auditions). It has been great getting answers to things I have been wondering about in my solo projects and not had anyone to ask, from experts who I otherwise would not have access to."

"In-person meetings are far too cost prohibitive. The hotels are overly expensive and AAS membership grants you very little other than reduced registration. This strongly affects smaller colleges and institutions, and limits the participation of students who arguably would benefit the most."

"Although I have for 40+ years (Yikes!) enjoyed the in-person meetings, I strongly believe that the impacts of climate change are so severe that these must be curtailed. Much may be lost if the in-person meetings are ended, but -- well -- is the Earth truly facing a climate catastrophe or not?"

"I think virtual meetings are a very nice alternative, especially for people who may have social anxiety, as it still allows them to participate. As a community, we should welcome change and inclusivity, and should therefore try to include more virtual meetings. I understand the need/desire for in-person meetings as well, but sometimes they are less accessible for everyone. Therefore, I believe, both to reduce carbon footprint, and be more inclusive, that we should have virtual meetings at least every third meeting."

"For me, virtual conferences and talks have enabled a lot of connections that would have been impossible given my personal/family ability to travel -- even though I have abundant research support funds. While there are obvious downsides to virtual connections, I've been very grateful for those opportunities."

"Hybrid meetings are the worst of all worlds -- virtual conferences are fantastic, as long as there's a way to support conversations stemming from talks at sessions (Slido and Slack are the best for this that I've seen at meetings). Gather.town also helps recreate some of the informal interactions that occur at meetings in a virtual setting."

"Obviously a virtual meeting is not as much fun or beneficial as the immersive in-person experience, but it did save me an enormous amount of money. I definitely want to attend more virtual conferences as long as the experience can be as fully participatory as possible."

"Those who cannot or choose not to attend conferences in person should not impose their circumstances on the rest of us. Give them alternatives and let them pay for those alternatives without burdening in-person participants."

"I already had strong feelings about attending due to the carbon impacts, but the pandemic ended up just clinching it for me. We've seen that virtual meetings are possible and they can be positive experiences. They include people with disabilities in ways they were excluded in the past, they reduce our carbon impact, and going forward it can help protect the most vulnerable in our society as we continue to deal with Covid and its variants for years to come. (The flu pandemic of 1918 never actually ended, and this virus is far more contagious.) The opposition to virtual meetings is anathema to me given all these benefits."

"As a community college professor teaching over spring, summer, and fall, it is very difficult for me to have time to travel to a meeting. The virtual opportunities over the last few years has allowed me to participate and keep in the loop about new research. Being able to watch the talks that I could not attend during the virtual meetings (i.e., which were during my classes), was great too. I absolutely would NOT have been able to do so without the virtual option."

"My observations is that the senior folk are less likely to come to in-person these days, which is not good for the younger people, who cannot network with the senior folks if they aren't there. This is a really important aspect of nurturing the next generation and we need to encourage in-person participation."

"The largest impact I have is due work-related travel. Now that the pandemic has taught us that we can cope with not always being present at conferences, I would strongly advocate in favor of a permanent hybrid status of conferences and other meetings. This would not only mitigate our footprint, but it would also help people who deal with disabilities (their own or in the family), caregivers and people who have limited economic resources to be really included in the discourse. Fighting climate change can foster inclusiveness."

"I believe that reducing our carbon footprint is incredibly important, but I am concerned about the recent push to move conferences to online-only venues. Senior scientists have benefited immensely from the networking and career advancing opportunities that come from meeting colleagues at in-person conferences, and it is unfair to students and junior scientists to not give them similar opportunities."

"I've actually been able to attend virtually conferences I wouldn't have otherwise been able to go to. I think the virtual option has made the meetings accessible in a way that was impossible before."

"Experiment with new forms of collaboration. This "virtual meetings" experience is still pretty new for us, and new tools (e.g. VR) have not fully caught on. We need some experimentation in meeting format to find the best mix that allows networking as well as traditional talks."

"Lastly, it is very easy to fall into the trap of online experience is just not comparable to in person, if we never invest resources towards improving it. It is high time we as a scientific community take into account the evident climate change issue and model our behavior for the rest of the society."

"The fully virtual AAS worked well for attending talks. What was completely lacking was the ability to interact one on one."

"I think the successful virtual meetings I have attended [...] have had successful interactions because there are a few key people who interacted a lot and thus helped drive / draw out interaction in other people."

"Attending a virtual meeting while at the home institution is problematic, since regular work meetings continue and I feel compelled to attend."

Re: Virtual Meetings: "More participants need to buy into the idea for it to work."

"When I travel to them, the AAS is the highest priority and I can be fully engaged. So, in some ways, a barrier to "attending" virtually is that I, and I have noticed my colleagues behaving in the same way, place a lower priority on actually engaging and being present. In my opinion, this reduces the overall quality of the meeting for everyone."

"The meeting was local for me but I was also recovering from recent surgery. I did attend some of the meeting in person as I was able, but it was very helpful to be able to attend a lot of it remotely when I didn't have the strength to go in person."

"Winter 2023 AAS. The interface was really quite good, and I found myself thinking that, if I were to attend a hybrid meeting in person, I would hole up in my hotel room and attend the talks (rather than rush around from room to room), then go to the meeting for social interactions, posters, etc."

"The ambiance and aura are not there in the virtual meeting."

"My engagement is so much poorer for virtual that they're useless. My fault, not the conferences."

"[Virtual meetings] also make it much easier for the audience (from senior scientists to students) to be distracted and/or working simultaneously."

"AAS meetings--more than any other astronomy meeting--are more than just the science, they are for the people and the unforeseen opportunities."

"Science is a human endeavor that thrives on human interaction."

"A mix of virtual and in-person seems like a possible good solution. No solution will satisfy everyone! The challenge with doing both is avoiding the trap of the virtual meeting being considered less utility, less impactful, and skipped by those who can afford to attend in person."

"Developing hybrid/virtual conferences is building the way to a better future and it's important that the AAS leadership encourages especially the more conservative part of the community to move in that direction."

"Instead of hosting two fully hybrid meetings per year, host 1 in-person focussed (virtual viewing only, which would be very low cost) and the other pure virtual. Not be complicit in organizing meetings that have a high carbon footprint per participant junkets, such as the upcoming Extreme Exoplanets meeting in Christchurch NZ."

"Meetings that are 100% in person or 100% online are easier to participate in and run than hybrid meetings, because you only need to focus on one mode of communication and one audience. There will always be people who can't travel because of limited funding, health concerns, or personal responsibilities, so meetings that do not require travel have a big advantage in terms of ideas and expertise because they have access to a broader range of potential speakers."

"I like the idea of assigned (possibly randomly) small group social events, like 30 min coffee breaks with a small (5--10 people) group and a discussion topic."

"Please please please find a way to facilitate interactions with iPoster presenters!"

"I would recommend checking out the work done by the future of meetings group: https://thefutureofmeetings.wordpress.com/"

"Meetings really do have the potential to be COVID superspreader events and this should definitely be taken into account in planning right from the beginning."

"Ask registrants if they have attended an in-person meeting in the last year and make that a negative impact on registration for the meeting if they answer yes."

"I think the buildings where AAS meetings should be held should be very insulated and energy efficient, and share this information with its members. It should also provide receptions that are only vegetarian and or vegan and never use single plastics."

"Virtual conferences also prevent mixing of generations as many virtual presentations end up being by early career scientists, while the mid-career and later career people end up playing a very passive role (if attending at all)."

"It's important to consider the (differential) impact that AAS's actions addressing climate change will have on different career stages, identities and backgrounds, alongside implementing climate-friendly policies."

"If there were only one winter in-person AAS meeting every 4 years, then it would seem like more of a special event to be prioritized, than if it's every year and there's always next year. That could be staggered with one summer in-person AAS meeting every 4 years. And each could be virtual for 3/4 years."

"I think the AAS should consider alternating between virtual and in-person meetings. This would force an attempt to truly make the virtual meetings work. After several cycles we would be able to decide if it worked or if we preferred either virtual or in-person so much we wanted to go fully (or more fully) to one or the other."

"Recycle/reuse name tags (SPIE does this). Hold conferences in cities with recycling and composting programs and public transit to/from conference and to/from airport, continue to use only compostable single-use materials, encourage vendors to share non-plastic merch and virtual literature when possible, provide registration discounts or meal/drink tokens to local astronomers who are not flying in, provide other perks for attendees and vendors who comply with a number of eco-friendly tasks from a list (ex: no plastic or styrofoam freebies, didn't fly, has reusable water bottle, walking from hotel to conference, sharing a hotel, etc... make a bingo game of it, maybe)"

"The largest and most obvious step the AAS could take towards this goal is to reduce the frequency of general meetings to once per year (or once per 18 months? per 24 months?). To compensate for this, the AAS could assist its members in organizing more mid-sized regional conferences. This would reduce carbon-intensive long-distance travel by AAS members without eliminating opportunities for in-person interactions."

"In-person meetings should be rare (limit carbon footprint) and happen in the summer (limit COVID risk)."

"I enjoy having captions on with Zoom (I'm not hard of hearing but do have some problems processing speech) and find it makes everything more accessible to have captions as an option. It also is less physically exhausting or overstimulating to attend virtually, which meant I could attend a lot more talks and town halls than would have otherwise been possible."

"AAS is a huge, exhausting meeting in person. Virtual meetings seemed like only the exhausting parts (talks, posters) without any of the energizing benefits (personal connections, etc.)"

"There have been some topical meetings that have been very successful in the online format. In particular, there was a KITP workshop on the circumgalactic medium that had a really nice combination of plenary talks, targeted talks, workshops, and other things, with recordings of all of those being available. There was also a robust Slack group with a high level of engagement that had some very clever structures in place - aligning Slack channels to go along with meeting sessions, in particular - and strong encouragement for people to create and post short YouTube videos on their recent papers/research, topics of interest, etc. A lot of it is asynchronous and that helped with engagement quite a bit. I also participated in a meeting that had asynchronous talks, but synchronous Q&A/discussion for the talks (which was recorded, and where people who couldn't attend could submit questions for consideration ahead of time). Auto-transcripting was used as well, and made available on YouTube along with the video so people could skim through and find the things they cared about. That allowed the synchronous session to be relatively time-compressed, and people who missed the synchronous session could keep up on what was going on."

"The best virtual meeting I attended had pre-recorded talks, so all of the online time was spent in discussion. That was by far a better use of time than watching talks online. That said, it also requires expert facilitation. I don't think just a room full of zoom participants is effective."

"Making a social hour with GatherTown or similar so that people can "run into" other people would really be more fun."

"I can't really buy a carbon offset with my grant, but the meeting could buy one and include it in the base price."

"Panel discussions among presenters in a special session make the virtual meeting more engaging– it's the "having a conversation" feel that's important to transfer from in-person to virtual meetings."

"Training for using the Meeting's Slack space for participants and vendors. The way that fully-virtual Meetings' Slack spaces are currently set up is extremely distracting because people with no training and/or sense of how to responsibly use Slack (or other forms of social media) are too easily able to, for example, tag literally everyone at the meeting and force notifications, at literally any time of day (i.e., during talks, meetings, etc.)."

"AAS could have virtual topical meetings throughout the year."

"Closed captioning of talks, for people with hearing disabilities. Questions in the chat, read by a moderator -- wonderful for increasing inclusion, and also wonderful for people with hearing disabilities. Some way to increase networking, and opportunities to talk with other researchers, without feeling like a crasher in a private club."

"Ask A Senior Astronomer" or "Ask An Undergrad" networking hours to facilitate cross-generational connections and information transfer?

"Chat-roulette-style social activities, themed gatherings."

"Support and design the meeting for virtual first– in person secondary only. Flip the narrative here."

"Should consider parallel, smaller meetings for local people to gather. This would ensure in-person participation and reduce travel."

"I would recommend that virtual meetings get rid of things that do not work, like the vendor booths. Nobody came to those. Maybe replace them with allowing each vendor a canned video that people could watch and put in comments/questions instead."

"Staffing a virtual exhibit hall booth is a rather strange experience and mostly seemed not terribly useful. It's a good idea but I'm not sure it really works in practice. That said, I'm not sure I have a good solution. Perhaps just a matter of treating it less as a live booth experience and more as an advertising and 'message board' kind of area, where attendees can also catch people's attention for a chat in relatively short order."

"Are you talking to the DPS committee that investigated ways to make meetings more inclusive? That committee did A LOT of work investigating pros/cons of virtual meetings."

"For hybrid meetings, each room should have two screens: one that shows what the presenter is wanting the audience to see [...] and another screen to show the remote audience members. Remote viewers should be encouraged to turn on their video cameras, especially when asking a question or certainly presenting / answering a question. There should also be ready break-out rooms for remote participants to use in private."

"If the virtual option wanted to produce a specific session tailored directly to students attending virtually, that could be an interesting way of helping students attend, feel like they are part of the AAS community, and network with other students around the country/world."

"Providing presenters and participants tips about virtual presentations is helpful, for example: Make sure you have an appropriate background or set a virtual background. Have the camera at your eye level and at an arm's length distance. Minimize background noise. There are many Youtube videos that provide useful tips."

"For meetings with a live component, being able to see the other participants is important - so having as many people as possible turn on their cameras (zoom or similar), or for a hybrid meeting, having a camera that is also focused on the audience, not just the speaker."

"My favorite thing was the random button to go through the posters."

"I stopped attending AAS meetings when I was a postdoc because I felt they were no longer contributing to my career. The talks are too short to really present anything, the makeup of the sessions is too random, and I noticed that every year there were more and more undergrads and fewer and fewer professors in attendance. That meant I had no one to network with. It became cheaper and more efficient to only attend the conferences in my field. The way the AAS is set up now, I feel that it's a good opportunity for undergrads and graduate students to get a feel for astronomy and get excited by the sheer size of the event. But I'm not sure it's still relevant for postdocs and professors."

"SDSS virtual meeting in 2020 worked surprisingly well. Sessions were shorter, and shifted in time from day to day so that individuals spread across time zones could still participate."

"The one feature that needed work was that it was almost impossible for in person people to mingle with online attendees (both in terms of the timing of sessions being tight, and infrastructure -- e.g., there could be a space at the conference site set aside for in person attendees to log on to Gathertown or equivalent)."

"Would having these meetings in the northeast where train rides to the venue would be feasible? Could there be incentives associated with attendees selecting less carbon intensive modes of transportation to the meetings?"

"Virtual meetings would be okay, but the cost is high if only a few talks are of interest. You may consider another society benefit and give out codes to attend 3 (or 5 or whatever) sessions of a virtual meeting at no cost. Or give codes to attend virtually for a day in exchange for helping with sorting abstracts. Maybe just save the last one for grad students and/or post docs though."

"Archiving talks and online Q&A (and setting up opportunities for asynchronous communication with the speakers) for virtual participants seems to be the best option for "hybrid" participation. This also reduces the resource load on the AV teams by lowering emphasis on "live" transmission."

"1) AAS should notify the organizations which are part of AAS to not interrupt with other work during the meeting, this letter should also be sent to astro departments. Just because people aren't physically at the meeting doesn't mean they don't need the same level of focus. 2) Provide at-home child support funds for parents so they can focus on the meeting. 3) Limit capacity for in person meetings. AAS is bleeding money on larger venues."

"I do like the idea for virtual meetings of assigned (possibly randomly) small group social events, like 30 min coffee breaks with a small (5--10 people) group and a discussion topic."

"I attended an AGU meeting where they provided several electric scooters for those who needed them on a daily basis. That was a big help."

"If hybrid meetings are going to be the expectation going forward [..], then I suggest the AAS stand up a "hybrid meeting team" that is responsible for the virtual component. It is not reasonable to ask the local organizers to basically organize two separate meetings at the same time, and the virtual component is really hard to do well. Having a single team that does this part and can preserve institutional memory might work better."

"I have attended many successful virtual meetings and it is a format that requires a lot of advanced preparation and professionalism. For example, the National Academy of Sciences is very experienced with these virtual meetings and these gatherings are always on time, have a clear agenda and the speakers are carefully selected, and in many cases they distribute the material presented."

"The functions of invited and contributed papers plus discussion, town halls, posters plus interaction with authors are better matched to virtual than to in person. Virtual workshops are as good as in person, and better than mixed. The part of FTF that virtual cannot replace is casual interaction, drifting in the crowd, etc. So, put all the above in a virtual world. Schedule FTF as short (few days) events devoted entirely to "other" especially "walking around", probably with theme centers and activities primarily to promote interaction, including cross-discipline, cross-cultural, cross- age group. To suppress the high cost of mega-facilities, plan this as relatively frequent smaller and more scattered venues. As in, well, I have some time in April, where is the AAS Get-together nearest to me?"

Examples of Successful Virtual and Hybrid Meetings

"I liked how the "Cosmology at Home" meeting was organized. [...] Participants worked together on small projects during the meeting too due to "unconference sessions" that were carefully planned."

"For each good meeting, I had multiple screens open, with multiple things happening. The presentation was going, but on another screen were questions, comments, etc. A colleague of the presenter was online clarifying details. It also gives the audience a chance to participate."

"The Kavli Fundamentals of Gaseous Halos workshop, which was very different than an AAS meeting.  Each day had only a half day with some planned talks but driven by conversation groups, tutorials, and panel discussions."

"One of the things I liked about an EAS meeting I attended (2020?) was that if you left the platform open, you could hear an organizer announce through the system that it was time to gather for the next  session -- so suddenly I heard a voice out of my computer saying, "All right, everyone, grab your coffee, we'll be starting the next round of talks in two minutes." It startled me the first time, but I came to really appreciate it -- it raised my anticipation and made me feel that I was participating in a group activity, even if I couldn't see anyone."

"I liked one small virtual meeting I attended in 2020, because it was over several afternoons (so not all day) and it was the only time people attended the social aspect."

"Excellent audio quality (microphones hanging from ceilings), excellent visual quality, opportunities to contribute asynchronously before and after the meeting, virtual participants are prioritized in the Q&A session."

"The "Preventing Harassment in Science" virtual workshop hosted by LPI in 2020 was one of the most interactive virtual meetings I have attended. I believe this was largely due to the efforts of the speakers, who included interactive components in every single session of the meeting. The interactive sessions ranged from Q&A, online collaboration tools (idea boards), zoom chat interactions, etc. https://www.hou.usra.edu/meetings/anti-harassment2020/program/ "

"The SAZERAC virtual meetings [..,] as half-day affairs, they put less pressure on our scheduling. I also appreciated how they accumulate questions on-the-fly using a dedicated slack channel and then refer to them afterwards. It's much better than raising hands and not getting called on by the moderator."

"SAZERAC meetings -- they are designed to be virtual from the ground up. All participants join on slack, which is where all questions are asked and often answered in threads."

"The SAZERAC (high-z) meetings were great. There was a lot of flexibility for giving live or recorded talks, one could follow along on zoom or youtube, and having all questions asked and answered in slack (even the ones that were discussed live over zoom) made it easy to engage and follow discussions asynchronously."

"LSST Dark Energy Science Collaboration runs excellent virtual meetings for 200 active members. Mixes plenary, parallel, social events, and breaks. The best parallel sessions use breakout rooms for small-group discussions."

"Cool Stars 20.5 - Astronomers for Planet Earth symposium: Pre-recorded talks and all the live time was spent on panels, discussion, and other interactive formats."

"I have liked the more interactive Zoom-like sessions with chat options IF there was a good moderator and if there were no barriers for presenters to connect - pre-recorded talks have helped with that and with addressing bandwidth limitations for some meetings."

"Lilly Online Teaching Conferences are quite good. The online schedule is quite navigable. They have lots of ways to encourage people to explore the virtual space: places to post photos, chime into a discussion thread, send a "hello" message to "connect" with other attendees. They sometimes have a low key contest going on all week to see who can make the most contributions in the aforementioned ways. The high scorers are listed on a leaderboard. They also have a prize raffle drawing in the last session where they give away books and other desirable items. Everyone gets a chance to vote beforehand which prizes you would or would not want, to ensure you don't end up with junk you don't want. They read the winners' names in the last session. Must be present to win. It encourages folks to attend the last session."

"I attended the virtual APS April Meeting in 2020. One of the best things about it, I thought, was that the session streams were DVR-like: that is, a session could be watched live but it also could be paused, rewound, watched with a small or large delay, and played back at 1.5x speed.

Personally, I found that very helpful to attend all of the sessions and talks I wanted, even when some of those sessions were directly conflicting, while still feeling engaged in the meeting."

"In December 2020, the virtual conference "Five Years after HL Tau" organized by ESO and NRAO set the standard for me for such events. There were daily 4 hour synchronous sessions over zoom, with invited review talks, short Q&A, and two longer panel discussions. In addition, there were asynchronous pre-recorded talks and posters, with slack channels. Registration was free. There were ~500 participants, and everything worked very smoothly."

"Heliophysics2050. It is the only virtual workshop I have seen that managed to replicate the informal conversations and discussions common in in-person meetings. I think that this was likely a combination of two factors: (1) the workshop was about discussing what we wanted to promote for the Heliophysics Decadal Survey, so it was primarily ABOUT the discussion, not presenting research although that did happen and (2) people USED the Slack channel very heavily. I have never seen so much participation in online chat channels at a conference before or since. It felt like the whole attendance list was engaged with the platform, and there were multiple threads going in every channel, and often in the chat during panels/presentations. Even though I was online physically, I felt very connected to the conference and all those attending. This has not been true of many other larger and more recurring virtual or hybrid conferences I've attended. Unfortunately, I'm not sure it can be recreated, as it seemed to depend heavily on the fact that the conference had a targeted purpose and we were all attending with the intention to engage in that manner. That said, Metascience 2021 also managed to capture some of that activity and connectedness, to an extent. I do think it also helped that the various Slack channels were constantly talked about and advertised in the talks, so people knew where to find a particular discussion or who to reach out with a direct message about a particular topic."

"Remote Boston-Area Exoplanet Science Meeting in 2021. [...] There were short sessions of talks with frequent breaks and with a networking session that has led to fruitful interactions since. We were placed in breakout rooms of 3-4 people, went around and described our research, then asked each other questions about their work and got to know each other."

"Sagan Summer Workshop 2021 I liked the interactive nature of the speakers using polls and having hands-on data sessions. I also liked that talks were pre-recorded and available before the workshop so I could digest the material and ask better questions."

"Using slack channels has been an excellent way for discussion in smaller conferences for individualized sessions. Also, having the slack channel open before the meeting and set up for the individual sessions was really helpful."